\documentclass[aps,showpacs,twocolumn,10pt,pre]{revtex4-1}
\usepackage{amsfonts}
\usepackage{amssymb}
\usepackage{graphicx}

\usepackage{indentfirst}

\begin{document}
\title{Delayed double ionization as a signature of Hamiltonian chaos}

\author{F. Mauger$^1$, A. Kamor$^2$, C. Chandre$^1$, T. Uzer$^2$}
\affiliation{$^1$ Centre de Physique Th\'eorique, CNRS -- Aix-Marseille Universit\'e, Campus de Luminy, case 907, F-13288 Marseille cedex 09, France\\  $^2$ School of Physics, Georgia Institute of Technology, Atlanta, GA 30332-0430, USA}
%\date{\today}

\begin{abstract}
We analyze the dynamical processes behind delayed double ionization of atoms subjected to strong laser pulses. Using reduced models, we show that these processes are a signature of Hamiltonian chaos which results from the competition between the laser field and the Coulomb attraction to the nucleus. In particular, we exhibit the paramount role of the unstable manifold of selected periodic orbits which lead to a delay in these double ionizations. Among delayed double ionizations, we consider the case of ``Recollision Excitation with Subsequent Ionization'' (RESI) and, as a hallmark of this mechanism, we predict oscillations in the ratio of RESI to double ionization yields versus laser intensity. We discuss the significance of the dimensionality of the reduced models for the analysis of the dynamical processes behind delayed double ionization.
\end{abstract}

\maketitle

%%%%%%%%%%%%%%%%%%%%%%%%%%%%%%%%%%%%%%%%%%%%%%%%%%%%%%%%%%%%%%%%%%%%%%%%%%%%%%%%%%%%%%%%%%%%%%%%%%%%%%%%%%%%%%%%
%%                                          Introduction--Motivations                                         %%
%%%%%%%%%%%%%%%%%%%%%%%%%%%%%%%%%%%%%%%%%%%%%%%%%%%%%%%%%%%%%%%%%%%%%%%%%%%%%%%%%%%%%%%%%%%%%%%%%%%%%%%%%%%%%%%%

\section{Introduction} \label{sec:Intro}

The dynamical processes by which atoms lose their electrons when subjected to strong laser pulses usually fall into two main categories~\cite{Beck08}: sequential double ionization (SDI) in which electrons are removed one after the other, and a less trivial class of processes coined nonsequential double ionization (NSDI). In NSDI processes, the Coulomb interaction between two electrons plays an essential role. A particularly relevant example is afforded by the recollision scenario in NSDI~\cite{Cork93,Scha93} in which a preionized electron returns to the ion core to excite a bound electron. This excitation may lead to a double ionization if the energy transfer from the preionized electron to the bound electron is sufficient to overcome the Coulomb attraction to the nucleus, while the preionized electron keeps enough kinetic energy to remain ionized. This so-called ``direct impact ionization''~\cite{Rude04} is often invoked to explain the additional amount of double ionization at intermediate values of laser intensities. However the dynamics exhibits a much richer variety of double ionization processes. It has been observed that, sometimes, the recollision excites the parent ion which is ionized by the field after some delay, ranging from a quarter up to several laser cycles. These processes mimic a sequential double ionization process even though they fall into the NSDI category since the electron-electron interaction plays a role. This alternative and less straightforward road to NSDI is called Recollision Excitation with Subsequent Ionization (or RESI for short~\cite{Feue01,Rude04,DeJe04_1,DeJe04_2}). These processes are by no means rare and unimportant. For instance, it was shown that RESI plays a central role in High Harmonic Generation~\cite{Sand99} which is a way to generate short X-ray pulses. RESI processes belong to a category of double ionization involving a delay in the second ionization, and this category is referred to as delayed double ionization. The precise mechanism responsible for delayed double ionization, is still a subject of study from the theoretical and experimental points of view~\cite{Figu11}. Experimentally, the signature of RESI can be found in momentum distributions: The two lateral peaks in the bimodal distribution are attributed to direct impact ionization whereas RESI contributes to the central part around zero ion momentum~\cite{Webe00_2,Mosh00,Feue01,Rude04,Ruiz08}. It also turns out that direct impact ionization is expected to be triggered when the electric field is close to zero whereas RESI is expected when the electric field is maximum. 

\begin{figure}
	\centering
		\includegraphics[width = \linewidth]{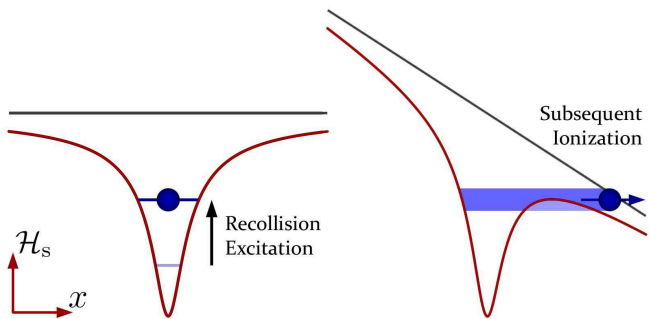}
	\caption{\label{fig:RESImechanism}
	(color online) Schematic representation of the conventional mechanism for RESI using a tunneling argument in the adiabatic approximation. Left panel: After a recollision with a preionized electron at a zero of the field, the inner electron is excited, but not ionized. Right panel: Later on, at an extremum of the field, the potential barrier is lowered by the laser field, and the excited electron tunnels. On each panel, the curves represent the effective potential (soft Coulomb potential plus the laser), the dark lines label the laser excitation and the horizontal lines represent the energy levels for the electron.}
\end{figure}

There is still some debate on the mechanisms of delayed double ionization~\cite{Ivan05,Figu11}. The conventional scenario for RESI is based on a tunneling argument in the adiabatic approximation. This scenario is depicted in Fig.~\ref{fig:RESImechanism}: After a recollision at a zero of the field, the remaining ion is in an excited state but not yet ionized. When the laser reaches an extremum, it creates a potential barrier through which the electron can tunnel. Therefore it raises a natural question if RESI is a purely quantum phenomenon. To counter this argument, RESI has been observed with classical calculations~\cite{Haan08,Shom09,Haan10}, for which all ionizations are over-the-barrier, casting some doubts on the validity of the tunneling argument in the adiabatic approximation. 

In this article we revisit this conventional picture of the RESI mechanism by analyzing the phase space mechanisms behind delayed double ionizations, of which RESI is a prime example. We show that after an electron is promoted to a specific region of phase space, it is trapped for some time before ionizing. This specific region is a chaotic one resulting from the competition (and resonances) of the Coulomb attraction with the laser field. At a given laser intensity, this thin chaotic region is organized by a single resonance through the hyperbolic periodic orbit it creates. The chaotic nature of RESI trajectories was already identified in Ref.~\cite{Sand99} through the ``deflection function''. The way an electron is ionized in the chaotic region is by following the unstable manifold of some selected periodic orbits. As a consequence, the properties of these periodic orbits (as a function of the laser intensity) drive the delayed double ionization yields. As a signature of these periodic orbits, we find oscillations in the RESI to double ionization yields as a function of the laser intensity in a range of laser frequency. Some of these results were announced in a recent Letter~\cite{Maug12_1}.

In this paper, we restrict our analysis to linearly polarized fields. Interesting experimental and early theoretical results suggest the importance of delayed double ionization with elliptic or circular polarization as well~\cite{Pfei11}.

In Sec.~\ref{sec:cm}, we describe the classical models we use for the analysis of delayed double ionization. We exhibit some relevant reduced models, and show that they are pivotal for the identification of the delayed double ionization mechanism. In Sec.~\ref{sec:ddi}, we present some peculiar features of RESI trajectories in phase space. Using the reduced models of Sec.~\ref{sec:cm}, we provide a detailed analysis of the mechanisms in phase space. We show the importance of the dimensionality of the reduced models. 

%%%%%%%%%%%%%%%%%%%%%%%%%%%%%%%%%%%%%%%%%%%%%%%%%%%%%%%%%%%%%%%%%%%%%%%%%%%%%%%%%%%%%%%%%%%%%%%%%%%%%%%%%%%%%%%%
%%                                                   Models                                                   %%
%%%%%%%%%%%%%%%%%%%%%%%%%%%%%%%%%%%%%%%%%%%%%%%%%%%%%%%%%%%%%%%%%%%%%%%%%%%%%%%%%%%%%%%%%%%%%%%%%%%%%%%%%%%%%%%%
\section{Classical models} \label{sec:cm}

Remarkably, classical mechanics is able to identify both sequential and nonsequential pathways leading to double ionization and reproduces the experimental and computational observations closely~\cite{Panf01,Ho05_1,Maug09}. This success was ascribed to the paramount role of the electron-electron interaction which is present both in quantum and classical models. One of the advantages of classical mechanics is to analyze these systems in phase space in a very detailed way. In Refs.~\cite{Maug09, Maug10} it was shown how invariant structures like periodic orbits rule double ionization processes in strong and short laser pulses. The methodology here is to follow the same modus operandi in order to provide insights into delayed double ionization. In this section we detail the various models which will be used later in the article. 

The parent model is a two active electron atom with soft-Coulomb potentials subjected to an intense and short linearly polarized laser pulse in the dipole approximation~\cite{Panf01,Ho05_1,Maug09}. The corresponding Hamiltonian reads:
\begin{widetext}
\begin{eqnarray}
  {\mathcal H} \left( {\bf x}_{1}, {\bf x}_{2}, {\bf p}_{1}, {\bf p}_{2}, t \right) & = &
      \frac{\left\|{\bf p}_{1}\right\|^{2}}{2} + \frac{\left\|{\bf p}_{2}\right\|^{2}}{2}
      - \frac{2}{\sqrt{\left\|{\bf x}_{1}\right\|^{2} + a^{2}}}
       - \frac{2}{\sqrt{\left\|{\bf x}_{2}\right\|^{2} + a^{2}}}
      + \frac{1}{\sqrt{\left\|{\bf x}_{1}-{\bf x}_{2}\right\|^{2} + b^{2}}}  \nonumber \\
     & & + \left( {\bf x}_{1} + {\bf x}_{2}\right) \cdot {\bf e}_{x} E_{0} f\left( t \right) \sin \omega t  \label{eq:FullHamiltonian}
\end{eqnarray}
\end{widetext}
where ${\bf x}_{i}=\left(x_{i},y_{i},z_{i}\right)$ is the position of the $i^{\rm th}$ electron ($i=1,2$) and ${\bf p}_{i}=\left(p_{x,i},p_{y,i},p_{z,i}\right)$ its canonically conjugate momentum. Here ${\bf x}$ and ${\bf p}$ belong to $\mathbb{R}^{d}$ for $d=1$, 2 or 3. The nucleus is assumed to be fixed at the origin because of its large mass compared to that of the electrons and the relative short duration of the laser pulse. The linearly polarized laser field (along the direction ${\bf e}_x$) is characterized by its amplitude $E_{0}$ and has a wavelength of 780 ($\omega=0.0584$) or 460 nm ($\omega=0.1 \ \mbox{a.u.}$) which is specified for each computation. Its envelope $f\left(t\right)$ is composed of a two laser cycle linear ramp up and a six laser cycle constant plateau. The constants $a$ and $b$ are the electron-nucleus and electron-electron softening parameters respectively and are chosen so as to be compatible with the ground state energy $\mathcal{E}_{g}$, defined as the sum of the first two ionization potentials~\cite{Java88,Panf01,Panf02,Haan94}. The same softening parameters are used independently of the dimensionality $d$ for a given atom.

The model~(\ref{eq:FullHamiltonian}) has two and a half degrees of freedom. In order to describe the dynamical processes in phase space, it is desirable to consider reduced models. These reduced models are described in what follows. We consider several kinds of reduced models: dimensionally reduced ones, and single electron models, or a combination of both. All of these models are used in a complementary way so as to get insights into the dynamical processes of delayed double ionizations.    

%% Dimensionally reduced models %%%%%%%%%%%%%%%%%%%%%%%%%%%%%%%%%%%%%%%%%%%%%%%%%%%%%%%%%%%%%%%%%%%%%%%%%%%%%%%%
\subsection{Dimensionally reduced two electron models}

From the dynamics associated with Hamiltonian~(\ref{eq:FullHamiltonian}), we notice that any plane containing the nucleus and the polarization axis is dynamically invariant. More specifically, starting with the initial condition $z_{1}=z_{2}=0$ and $p_{z,1}=p_{z,2}=0$, the two electrons remain in the $\left(x,y\right)$ plane. It is then possible to consider a dimensionally reduced model where the $z$-direction (and its canonical momentum) is left out. Such models are generally referred to as $2+2$ dimensional models, making reference to the dimension of the configuration space in which each electron moves. 

\begin{figure}
	\centering
		\includegraphics[width = \linewidth]{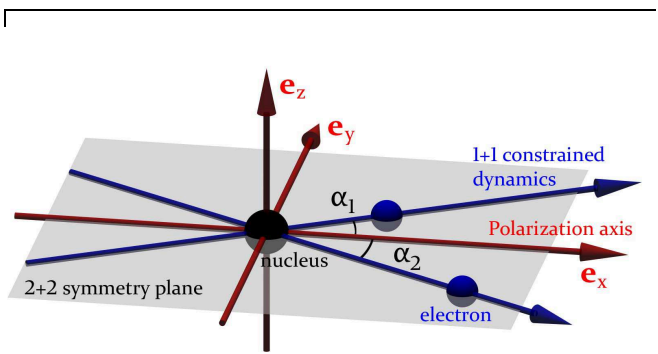}
	\caption{\label{fig:1Dmodels}
	(color online) Dimensional reduction for two active electron models. From the 2+2 dimensional model, we constrain the electrons to move along lines that form a constant angle with the polarization axis ${\bf e}_x$.}
\end{figure}

The $2+2$ dimensional model can further be reduced to a $1+1$ dimensional model by constraining each electron to move along a line which has a constant angle with the axis of polarization. We notice that, contrary to the reduction from $3+3$ to $2+2$ dimensions which can be carried out independently of the ellipticity of the field, the one from $2+2$ to $1+1$ can be considered only for linear polarization. The resulting constrained dynamics is obtained using, e.g., Dirac's theory of constrained Hamiltonian systems~\cite{Dira50} and the resulting Hamiltonian reads
\begin{widetext}
\begin{eqnarray*}
   {\mathcal H}_{22} \left(x_{1}, x_{2}, p_{1}, p_{2}, t \right) & = & \frac{p_{1}^{2}}{2} + \frac{p_{2}^{2}}{2} 
      - \frac{2}{\sqrt{x_1^{2} + a^{2}}} - \frac{2}{\sqrt{x_2^{2} + a^{2}}}
       + \frac{1}{\sqrt{(x_1- x_2)^{2} + 4 x_{1} x_{2} \sin^{2} \frac{\alpha_{1} - \alpha_{2}}{2} + b^{2}}} \\
      & & + \left( x_{1} \cos \alpha_{1} + x_{2} \cos \alpha_{2} \right) E_{0} f \left( t \right) \sin \omega t,
\end{eqnarray*}
\end{widetext}
where $\alpha_{i}$ is the angle with the axis of polarization (see Fig.~\ref{fig:1Dmodels}). Two special cases have been studied in the literature. They correspond to $\alpha_{1}=\alpha_{2}=0$, where the electrons are aligned with the axis of polarization~\cite{Java88,Panf01,Panf02,Haan94}, and $\alpha_{1}=-\alpha_{2}=\pi/6$~\cite{Prau05, Eckh06}. In the first case, the reduced Hamiltonian reads:
\begin{eqnarray}
   {\mathcal H}_{11} \left( x_1, x_2, p_{1}, p_{2}, t \right) & = & \frac{p_{1}^{2}}{2} + \frac{p_{2}^{2}}{2}
      - \frac{2}{\sqrt{x_1^{2} + a^{2}}} - \frac{2}{\sqrt{x_2^{2} + a^{2}}}  \nonumber \\
      & & + \frac{1}{\sqrt{(x_1- x_2)^{2} + b^{2}}} \nonumber \\
      & & + \left( x_1 + x_2\right) E_0 f(t) \sin \omega t. \label{eq:Ham2e}
\end{eqnarray}
We note that the reduced dynamics corresponds to a symmetry of original $2+2$ model. In the second case, the reduced Hamiltonian reads:
\begin{eqnarray}
   {\mathcal H}_{11e} \left( x_1, x_2, p_{1}, p_{2}, t \right) & = & \frac{p_{1}^{2}}{2} + \frac{p_{2}^{2}}{2}
      - \frac{2}{\sqrt{x_1^{2} + a^{2}}} - \frac{2}{\sqrt{x_2^{2} + a^{2}}}  \nonumber \\
      & & + \frac{1}{\sqrt{(x_1- x_2)^{2} + x_{1} x_{2} + b^{2}}} \nonumber \\
      & & + \left( x_1 + x_2\right) \tilde{E}_0 f(t) \sin \omega t, \label{eq:Eckh:Ham2e}
\end{eqnarray}
where $\tilde{E}_{0}=E_{0}\sqrt{3}/2=E_{0}\cos\pm \pi/6$ is the scaled peak field amplitude. 
Here the lines along which each electron is moving are not conserved by the flow of Hamiltonian~(\ref{eq:FullHamiltonian}). The main reason why Hamiltonian~(\ref{eq:Eckh:Ham2e}) has been considered in the literature is to allow the escape of electrons simultaneously with symmetrical momenta. In the numerical calculations, we use the two models independently and show that the dynamical mechanisms that regulate delayed double ionization are qualitatively the same for both $1+1$ dimensional models. 

Although the $1+1$ dimensional reduction may seem as a crude approximation, it has shown to be surprisingly accurate in reproducing quantum and experimental results in double ionization qualitatively, such as the characteristic knee shape in double ionization yield as a function of the laser intensity. This is the simplest model looking at the correlated dynamics of two electrons. In this case, the role of the softening parameters is to allow the charged particles to pass through each other along the polarization axis, as they would have done in a transverse direction with a higher dimensional model. 

%% Reduced one electron models %%%%%%%%%%%%%%%%%%%%%%%%%%%%%%%%%%%%%%%%%%%%%%%%%%%%%%%%%%%%%%%%%%%%%%%%%%%%%%%%%
\subsection{Reduced one electron models} \label{sec:InnerElectronModel}

Although the two electrons are treated equally in the previous models, a careful analysis of the dynamics reveals that, given some initial conditions, their dynamics are qualitatively very different: typically one electron is quickly ionized (the outer electron) while the other one remains close to the core (the inner electron)~\cite{Maug09, Maug10}. This observation stands independently of the chosen dimension, and it is at the core of the three step scenario~\cite{Cork93,Scha93}. As a result, for each electron an effective reduced model can be crafted. When this outer electron is far away from the core the electron-electron interaction can be neglected. Therefore the effective Hamiltonian for the inner electron is built from Hamiltonians~(\ref{eq:FullHamiltonian}), (\ref{eq:Ham2e}) and~(\ref{eq:Eckh:Ham2e}):
$$
   {\mathcal H}_{\rm in} \left( {\bf x}, {\bf p}, t \right) = \frac{\left\|{\bf p}\right\|^{2}}{2} - \frac{2}{\sqrt{\left\|{\bf x}\right\|^{2}+a^{2}}} + x \tilde{E}_{0} f \left( t \right) \sin (\omega t+\phi_0),
$$
where $\tilde{E}_{0}=E_{0}\cos\alpha$ is the scaled laser amplitude in $1+1$ dimensional models~(\ref{eq:Ham2e}) and~(\ref{eq:Eckh:Ham2e}), and  $\tilde{E}_{0}=E_{0}$ for higher dimensional models. We notice that, up to a scaling of the field amplitude, the same models for the inner electron stand independently of the chosen angle $\alpha$. To simplify the comparison between the models, in all numerical results we drop the laser amplitude scaling parameter (i.e., we choose $\tilde{E}_{0}=E_{0}$). This model for the inner electron is only valid after the outer electron has left the nucleus (i.e., after the preionization at $t=t_0$ such that mod$\left(t_{0},2\pi\right)=\phi_{0}$ and outside of the recollision events). For convenience, we shift time by $t_0$ so that the inner electron model is valid for $t\geq 0$. Since most of the nonsequential double ionization processes happen in the plateau~\cite{Maug09}, the model for the inner electron is further simplified as
\begin{equation} \label{eq:He+}
   {\mathcal H}_{\rm in} \left( {\bf x},  {\bf p}, t \right) = \frac{\left\|{\bf p}\right\|^{2}}{2} - \frac{2}{\sqrt{\left\|{\bf x}\right\|^{2} + a^{2}}} + x E_{0}\sin(\omega t+\phi_0),
\end{equation}
where $\phi_0$ denotes the laser phase at which the inner electron model is started (e.g., after the preionization $t_{0}$ or after the end of the final recollision). In what follows, we use these reduced models to highlight the relevant mechanisms at play for the delayed double ionization. 

%%%%%%%%%%%%%%%%%%%%%%%%%%%%%%%%%%%%%%%%%%%%%%%%%%%%%%%%%%%%%%%%%%%%%%%%%%%%%%%%%%%%%%%%%%%%%%%%%%%%%%%%%%%%%%%%
%%                                  Mechanisms for delayed double ionization                                  %%
%%%%%%%%%%%%%%%%%%%%%%%%%%%%%%%%%%%%%%%%%%%%%%%%%%%%%%%%%%%%%%%%%%%%%%%%%%%%%%%%%%%%%%%%%%%%%%%%%%%%%%%%%%%%%%%%
\section{Mechanisms for delayed double ionization} \label{sec:ddi}

%% Delayed double ionization trajectories %%%%%%%%%%%%%%%%%%%%%%%%%%%%%%%%%%%%%%%%%%%%%%%%%%%%%%%%%%%%%%%%%%%%%%
\subsection{Delayed double ionization trajectories}

First we begin by analyzing some sample trajectories of the two electron Hamiltonian~(\ref{eq:FullHamiltonian}). For the display, we only consider the case $d=2$. Other dimensional models lead to a similar typology of trajectories and the main differences between the various dimensional models are analyzed in Sec.~\ref{sec:dim}. Among all the two-electron trajectories, we consider the ones which are doubly ionized at the end of the laser pulse, and which present a significant delay between the last time the preionized electron influences the core dynamics and the subsequent ionization. These trajectories will be referred to as delayed double ionizations. We use an energy criterion to identify if an electron has ionized or not~\cite{Panf02, Maug09, Maug10}. The energy~$\mathcal{E}$ of an electron is defined as the sum of its kinetic energy plus the (soft) Coulomb interaction with the nucleus:
$$
   \mathcal{E} = \frac{\left\|{\bf p}\right\|^{2}}{2} - \frac{2}{\sqrt{\left\|{\bf x}\right\|^{2}+a^{2}}}.
$$
An electron is considered ionized whenever its energy is positive, i.e., $\mathcal{E}>0$. As for the electron-electron interaction, there are many ways a preionized electron influences the dynamics of the inner electron. The most famous one in strong field physics is the recollision. A recollision is said to have occurred whenever the distance between the two electrons is smaller (or equivalently, whenever the Coulomb interaction between the two electrons is larger) than some threshold. Of course the definition of the threshold is somewhat arbitrary. However on inspection of a large sample of trajectories, it leads to the definition of a recollision where there is a significant exchange of energy between the two electrons. 

\begin{figure}
	\centering
		\includegraphics[width = \linewidth]{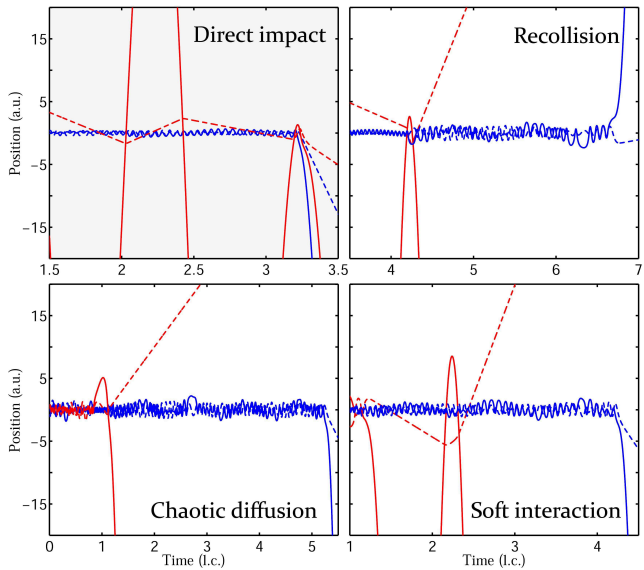}
	\caption{\label{fig:traj}
	(color online) Typical nonsequential double ionization trajectories for Hamiltonian~(\ref{eq:FullHamiltonian}) with $d=2$. The parameters of the atom are $a=b=1$, and the initial conditions have a microcanonical distribution on the energy surface $\mathcal{E}_{g}=-2.3 \ {\rm a.u.}$~\cite{Maug09}. The laser parameters are $I=3\times 10^{15} {\rm W}~ {\rm cm}^{-2}$ and $\omega=0.0584$~a.u. The top left panel corresponds to the conventional direct impact scenario while the other panels display the alternative delayed ionization routes (as indicated on the figure). Continuous (dashed) lines correspond to the dynamics in the polarization (transverse) direction.}
\end{figure}

We identify three main types of delayed double ionizations (to be specified later on) and give an illustration of them in Fig.~\ref{fig:traj}, and  compare them to the standard direct impact mechanism (top left panel). The common feature of all the delayed double ionizations shown in Fig.~\ref{fig:traj} is that the inner electron's role is very far from being a passive one, waiting for a recollision to ionize as in the standard picture of NSDI. In what follows we show that the inner electron is promoted to a specific chaotic region of phase space where it is trapped for some time (which might last up to several laser cycles). The way the inner electron is promoted to this region varies according to the specific trajectory and to the values of the laser parameters. It is a combination of several mechanisms: recollision, soft interaction and chaotic diffusion. The differences between these three types may seem very subtle at first glance. However, by looking closely at the trajectory of the inner electron right before being hit by the outer electron, we notice that the transfer of energy is very low for the soft interaction (lower right panel) while it is significant for the recollision (upper right panel). The chaotic diffusion (lower left panel) corresponds to the extreme case where there is no recollision during the plateau (note that during the brief excursion of the firstly ionized electron around $1$~laser cycles, the electron is not ionized). It means that the inner electron is promoted to the chaotic layer by the correlated dynamics of the two electrons during the ramp up, and no recollision is involved in the subsequent ionization (leaving out the ``R'' or RESI). Note that chaotic diffusion is not a marginal phenomenon among delayed ionization. All three types of delayed double ionization share the common featured that the inner electron is trapped in a chaotic region of phase space for some time before it ionizes. The difference between them is the way it reaches this region, i.e., through a (soft or strong) recollision or the correlated dynamics of the two electrons during the ramp up of the field. Among the three types, RESI trajectories, where a recollision is involved before ionization, are the most famous one. In this case, the significant time delay is compared to direct impact double ionization, where the subsequent ionization immediately follows the recollision (see the upper panels in Fig.~\ref{fig:traj}). This delay might last between a quarter up to several laser cycles. As expected, a typical delayed double ionization is the result of a cooperation between the three mechanisms, so even though we have tried to classify these trajectories into three categories, the boundaries between them are not sharp at all.    

%%%%%%%%%% Poincare sections %%%%%%%%%%%%%%%%%%%%%%%%%%%%%%%%%%%%%%%%%%%%%%%%%%%%%%%%%%%%%%%%%%%%%%%%%%%%%%%%%%%%%%%%%%%
\subsection{Poincar\'e sections} \label{sec:PS}

\begin{figure}
	\centering
		\includegraphics[width = \linewidth]{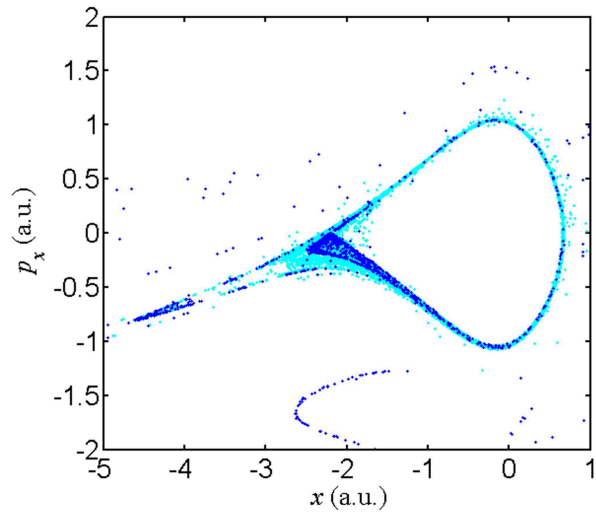}
	\caption{\label{fig:Real_RESI}
	(color online) Poincar\'e section (stroboscopic plot at the maxima of the field) of RESI trajectories after the last recollision of Hamiltonian(\ref{eq:Ham2e}) with $d=1$ . The parameters for the atom and the laser are the same as in Fig.~\ref{fig:traj}. For each trajectory, the first point on the section (so right after the last recollision) is plotted in light gray (blue online) while the following ones are in dark gray (blue).
	}
\end{figure}

In order to get deeper insights into the chaotic region in which the inner electron is promoted, we consider Poincar\'e sections of RESI trajectories of Hamiltonian~(\ref{eq:FullHamiltonian}) with $d=1$. We have chosen RESI because it constitutes a significant subset of delayed double ionization, and it is easier to detect numerically. In Fig.~\ref{fig:Real_RESI}, we display the positions of the electron which ionizes last at the successive maxima of the laser field after the last recollision. Looking at Fig.~\ref{fig:Real_RESI}, we note that RESI trajectories are restricted to specific parts of phase space. In particular, we note a central region, close to the nucleus where there is no record of RESI. This region is delimited by a thin layer where most of the inner electrons are promoted right after a recollision. In addition, we observe swirling patterns that extend far away from the nucleus. 

\begin{figure*}
	\centering
		\includegraphics[width = \linewidth]{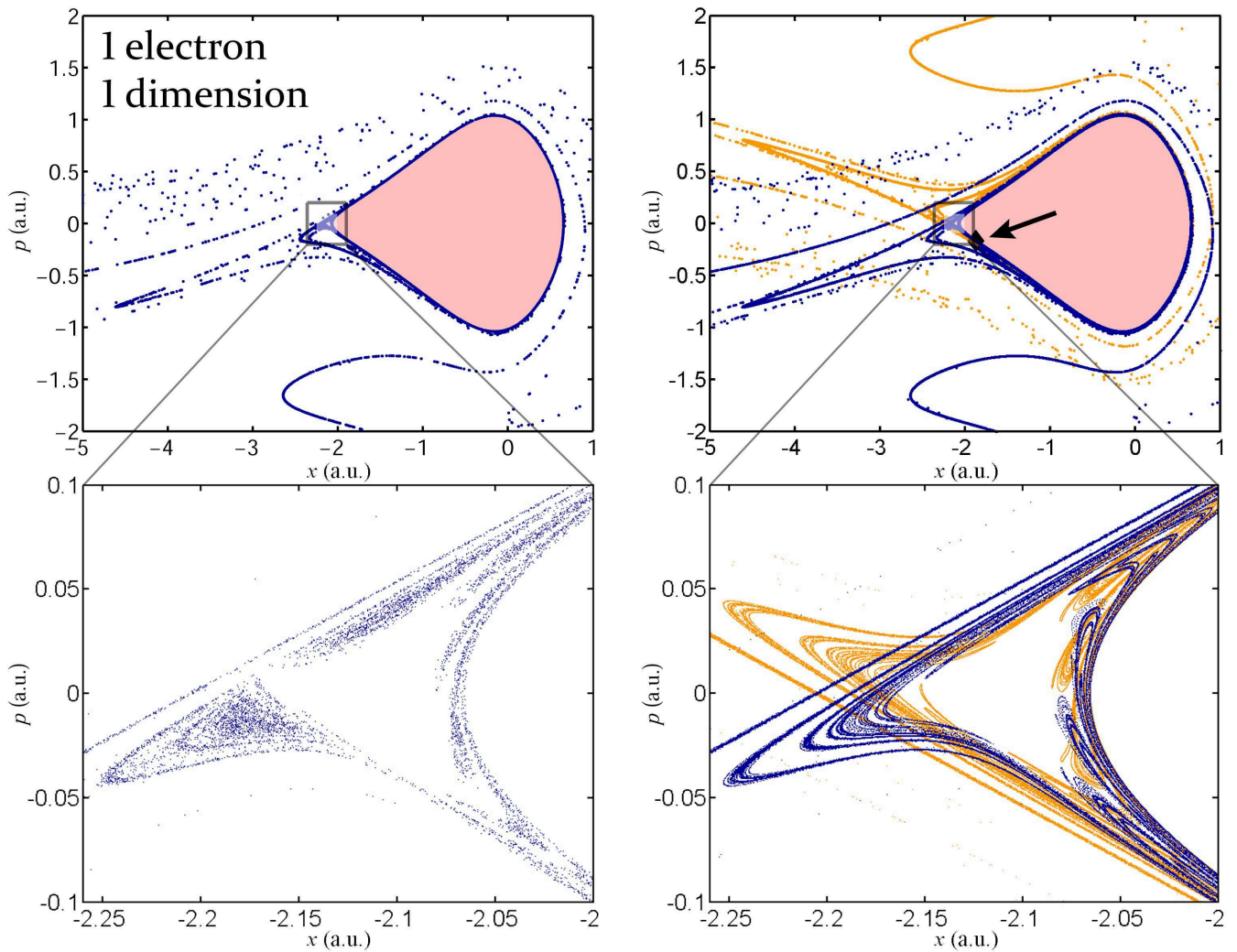}
	\caption{\label{fig:PSOS_RESI1}
	(color online) Left panels: Poincar\'e section (stroboscopic plots at the maxima of the laser field, i.e., $\phi\equiv\pi/2$) of RESI trajectories of the Hamiltonian~(\ref{eq:He+}) with $d=1$ detected from a large set of random initial conditions. Right panels: Poincar\'e section of the stable [light gray (orange online) dots] and unstable [dark gray (blue) dots] manifolds of the periodic orbit $\mathcal{O}_n^h$ of Hamiltonian~(\ref{eq:He+}) with $d=1$. The position of the periodic orbit $\mathcal{O}_n^h$ is indicated with a diamond (see the arrow) in the upper right panel. The light gray (pink) area in the upper panels represents the part of phase space from which the inner electron does not ionize. The lower panels focus are insets of the upper panels. The parameters for the atom and the laser are the same as in Fig.~\ref{fig:traj}.}
\end{figure*}

After a recollision, the inner electron is left with the nucleus and field interactions. We use Hamiltonian~(\ref{eq:He+}) for a one-dimensional one-electron model. In the left panels of Fig.~\ref{fig:PSOS_RESI1}, we represent Poincar\'e sections (stroboscopic plots at the maxima of the laser, i.e., $\phi\equiv \omega t+\phi_0 \mbox{ mod } 2\pi =\pi/2$) of all the delayed ionizations (i.e., ionizations which take more than one laser cycle). For clarity we left out the two first points on the Poincar\'e section for each trajectory (the missing points are located in the same part of phase space as the remaining ones, but they clutter the fine structures depicted on the figure). Note the close similarity between the Poincar\'e section of trajectories of the two-electron model~(\ref{eq:Ham2e}) in Fig.~\ref{fig:Real_RESI} and the ones of the one-electron model~(\ref{eq:He+}) in Fig.~\ref{fig:PSOS_RESI1} (upper left panel). In particular the swirling patterns in both figures are nearly identical.  

As shown in Figs.~\ref{fig:Real_RESI} and~\ref{fig:PSOS_RESI1}, RESIs are governed by the inner electron model~(\ref{eq:He+}) after the last recollision. More specifically, for one-dimensional models they correspond to a thin chaotic layer resulting from a resonance between the Coulomb attraction and the laser excitation. Since the one-dimensional one-electron model~(\ref{eq:He+}) captures the essential features of delayed ionization, this model is studied in more details in the next section. The motivation for studying one-dimensional models is twofold: First, the laser field drives the ionization dynamics along the polarization axis. Second, we find that this simplified dynamics forms the skeleton of higher-dimensional dynamics which, however, differs from the one-dimensional dynamics in significant ways, to be specified in Sec.~\ref{sec:dim}. 

%%%%%%%%%% One-dimensional one-electron dynamics %%%%%%%%%%%%%%%%%%%%%%%%%%%%%%%%%%%%%%%%%%%%%%%%%%%%%%%%%%%%%%%
\subsection{One-dimensional one-electron dynamics} \label{sec:InnerElectron:1DTimeDependent}

\begin{figure}
	\centering
		\includegraphics[width = \linewidth]{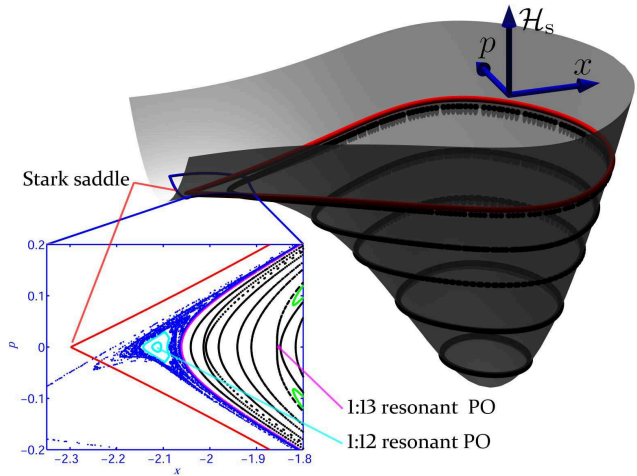}
	\caption{\label{fig:StaticFieldApproximation}
	(color online) Phase space of Hamiltonian~(\ref{eq:HamAdia}) for $d=1$ and for a laser intensity $I=3\times10^{15}~{\rm W}~{\rm cm}^{-2}$ (surface). The separatrix associated with the Stark saddle~\cite{ChaosAtomPhys} is represented as the outermost curve. For comparison, we display Poincar\'e sections (stroboscopic plot at the maximum of the laser field, dots) for Hamiltonian~(\ref{eq:He+}) with $d=1$. The inset shows a zoom of this Poincar\'e section close to the Stark saddle. We represent the outermost chaotic layer [outermost dots (blue online)], the boundary torus [outermost light gray (magenta) curve in the bounded region], the elliptic island around the $1$:$12$ resonant periodic orbit [light gray (cyan) curves] and the hyperbolic $1$:$13$ resonant periodic orbit.}
\end{figure}

In this section, we analyze the inner electron model given by Hamiltonian~(\ref{eq:He+}) with only one spatial dimension. It has one and a half degrees of freedom with no conserved quantity in the presence of a laser field ($E_0\not= 0$). Chaos is expected to develop, but only locally in phase space, depending on the parameter values of the laser field. To get a rough idea of the dynamics of Hamiltonian~(\ref{eq:He+}) on a short time scale compared to the period of the laser field, we consider a static field approximation where we freeze the laser field at peak field amplitude:
\begin{equation}
\label{eq:HamAdia}
   {\mathcal H}_{\rm s} \left( {\bf x}, {\bf p} \right) = \frac{\left\| {\bf p} \right\|^{2}}{2} 
      - \frac{2}{\sqrt{\left\| {\bf x} \right\|^{2}+a^{2}}} + E_{0} x.
   \end{equation}
Here we analyze this model for $d=1$. It exhibits a Stark saddle~\cite{ChaosAtomPhys} where the effective potential (Coulomb plus laser) is locally maximum and thus acts as a physical barrier to ionization. In phase space, the saddle has zero momentum and its position $x_{*}$ is a solution of $2x_{*}/\left(x_{*}^{2}+a^{2}\right)^{3/2}+E_{0}=0$. We denote ${\mathcal H}_{s}^{*}={\mathcal H}_{s}\left(x_{*},0\right)$ its energy. Since the time dependence has been removed, the system has one degree of freedom and thus it is integrable. In phase space, the curve ${\mathcal H}_{s} = {\mathcal H}_{s}^{*}$ is a separatrix (homoclinic connection) between the nonionizing trajectories and the ionizing ones (see the outermost curve in Fig.~\ref{fig:StaticFieldApproximation}). 

Taking into account the time-dependent potential, the dynamics of Hamiltonian~(\ref{eq:He+}) is slightly complicated by the possibility of chaotic trajectories near the saddle point. In fact, from KAM type arguments, it is possible to show the preservation of invariant tori (from the integrable dynamics described by ${\cal H}_{\rm s}$). We give an illustration of those tori in Fig.~\ref{fig:StaticFieldApproximation}. Looking closer at the dynamics, we notice some discrepancies between the static approximation and the time dependent dynamics. This is particularly the case close to the separatrix in Fig.~\ref{fig:StaticFieldApproximation}. The bound region of Hamiltonian~(\ref{eq:He+}) corresponds to a part of phase space, close to the nucleus, filled up with KAM tori. The outermost tori corresponds to a boundary torus [outermost light gray (magenta online) curve in the inset of Fig.~\ref{fig:StaticFieldApproximation}] outside which trajectories might be ionizing. This region is referred to as the unbound region, where the dynamics leads to ionization. Not all of the trajectories, though, ionize, as it can be seen for instance in a small elliptic island in the inset of Fig.~\ref{fig:StaticFieldApproximation}. We notice two main dynamical features of the unbound region: Close to the boundary torus, we observe a thin chaotic layer [outermost dots (blue online) in the inset of Fig.~\ref{fig:StaticFieldApproximation}]. Further away from the nucleus, the dynamics leads to rapid ionization (outermost white area in the figure).

In the bound region of the dynamics defined by Hamiltonian~(\ref{eq:He+}) with one dimension, not all tori survive from the integrable case without the laser field. The weakest ones are the resonant tori. They are characterized by $m\nu_{\rm f}=n\omega$, where $\nu_{\rm f}$ is the free field rotational frequency of the electron around the nucleus and $n, m\in\mathbb{N}^{*}$. They are later referred to as $m$:$n$ resonances. According to the Poincar\'e-Birkhoff theorem~\cite{RegAndStochastMotion}, when a resonant torus is broken at least two periodic orbits are created instead. From numerical simulations, we find two or four resonant periodic orbits depending on the intensity. 

\begin{figure}
	\centering
		\includegraphics[width = \linewidth]{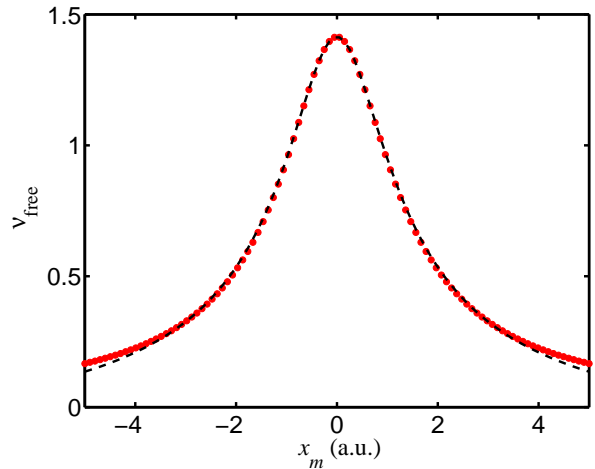}
	\caption{\label{fig:FreeFielFreq}
	(color online) Rotational frequency $\nu_{\rm f}$ of the electron around the nucleus (dots) as a function of the maximum position $x_{m}$. For comparison, we display the theoretical prediction (dashed curve) obtained as described in the text. The comparison is done without laser excitation and $a=1$.}
\end{figure}

For the softening parameters $a=b=1$, based on the observation that the rotational frequency depends almost linearly with energy, an analytical approximation for the action has been obtained in Ref.~\cite{Maug10}. The same formula can be generalized for other softening parameters, so that the approximate free field frequency $\nu_{\rm f}$ reads
\begin{equation} \label{eq:FreeFieldFrequency}
   \nu_{\rm f} = \gamma \left( \mathcal{E}-\mathcal{E}_{0}\right) + \omega_{0},
\end{equation}
where $\gamma$, $\mathcal{E}_{0}$ and $\omega_{0}$ are given by
$$
   \gamma = -\frac{9}{16}\sqrt{\frac{2}{a}}, \ \mathcal{E}_{0} =  - \frac{2}{a}, \ \omega_{0} = \sqrt{\frac{2}{a^{3}}}.
$$
The maximum position $x_{m}$ experienced by the electron is determined from $\mathcal{E}=-2/\sqrt{x_m^2+a^2}$. As a consequence we obtain the expression of the rotational frequency $\nu_{\rm f}$ as a function of $x_{m}$. In Fig.~\ref{fig:FreeFielFreq} we compare this prediction with the actual frequency, computed numerically. Note the very good agreement between the curves, even relatively far away from the nucleus.

\begin{figure}
	\centering
		\includegraphics[width = \linewidth]{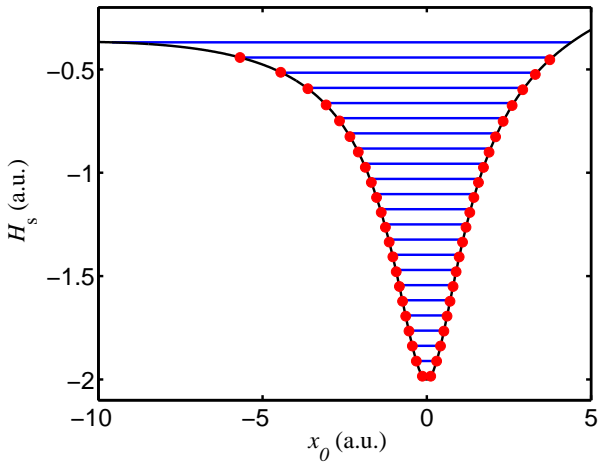}
	\caption{\label{fig:ResonantPO}
	(color online) Energy of the $1$:$n$ resonant periodic orbits at peak field amplitude as a function of their position $x_0$.  Horizontal lines label the predicted position for the resonant periodic orbits. Their actual position is depicted with dots. The parameters are $I_{0}=10^{13}\ {\rm W}~{\rm cm}^{-2}$, $\omega=0.0584$ and $a=1$.
	}
\end{figure}

Equation~(\ref{eq:FreeFieldFrequency}) combined with the resonance condition allows us to narrow down the regions of phase space in which to look for resonant periodic orbits. For the simplicity of the discussion we set the laser phase $\phi_{0}=\pi/2$, as most of our figures are drawn at peak field amplitude, such that $t=0$ corresponds to the peak field. In this case, for all $1$:$n$ resonances (the electron oscillates exactly $n$ times around the nucleus within one laser cycle), two resonant periodic orbits with zero initial momentum are present, one on each side of the nucleus. To predict their location, we use an adiabatic approximation: the total energy for the inner electron~${\mathcal H}_{\rm in}$, in the bound region, is almost constant in time. As a consequence, we replace the energy~${\mathcal E}$ with ${\mathcal H}_{\rm in}$ in Eq.~(\ref{eq:FreeFieldFrequency}) and the resonance condition becomes
\begin{equation} \label{eq:ResonantCondition}
   n \omega = \gamma \left( -\frac{2}{\sqrt{x_{0}^{2}+a^{2}}} + E_{0} x_{0} - {\mathcal E}_{0} \right) + \omega_{0},
\end{equation}
where $x_{0}$ is the position on the periodic orbit with zero momentum. Note that the implicit equation $x_{0}\left(E_{0},\omega,n\right)$ has in general two solutions, one positive and the other negative. They correspond to two periodic orbits with zero initial momentum located on each side of the nucleus. In Fig.~\ref{fig:ResonantPO} we compare the results given by this prediction with the actual location of the resonant orbits. The very good agreement with the prediction is worth mentioning. For a given softening parameter $a$, there is a finite number of primary resonances (and thus a finite number of resonant 1:$n$ orbits): From Eq.~(\ref{eq:ResonantCondition}) we deduce that the number of resonances $n_{\rm res}$ is equal to
\begin{equation} \label{eq:NbResonances}
   n_{\rm res} =\left\lfloor\sqrt{\frac{2}{a^{3}}} \frac{1}{\omega}\right\rfloor,
\end{equation}
where $\lfloor\cdot\rfloor$ is the floor function. For example, for $a=1$, we find $n_{\rm res}=24$ primary resonances. This number decreases as $a$ increases, i.e., as larger or equivalently more loosely bound atoms are modeled. 

For a given resonance index $n$, there are two periodic orbits with zero initial momentum, one on each side of the nucleus. We denote $\mathcal{O}_{n}^{-}$ the one located on the same side as the Stark saddle (see the inset of Fig.~\ref{fig:StaticFieldApproximation}) and $\mathcal{O}_{n}^{+}$ the other one. For the intensity range we have considered, ${\mathcal O}_{n}^{+}$ is found to be always elliptic while the stability of ${\mathcal O}_{n}^{-}$ depends on the order of the resonance. For odd $n$, ${\mathcal O}_{n}^{-}$ is hyperbolic for the whole range of intensities we have investigated. For even $n$, ${\mathcal O}_{n}^{-}$ is elliptic for intensities where it is in the bound region and, at higher intensities, after being released in the unbound region, it bifurcates into a hyperbolic periodic orbit. This bifurcation comes with the creation of an additional elliptic periodic orbit, through period doubling. For those resonances, in addition to the periodic orbits $\mathcal{O}_{n}^{\pm}$ we find two additional hyperbolic resonant periodic orbits, which we denote $\mathcal{O}_{n}^{h}$. They are connected through a symmetry of the system, such that we identify the two and restrict the analysis to only one of them (see Fig.~\ref{fig:PSOS_RESI1}, the diamond in the upper right panel).

As the laser intensity is increased, invariant tori are broken and the boundary torus between the bound and unbound region shrinks. This can also be seen in the static field approximation where this decrease is of the order of $E_0^{-1/2}$ (which can be seen from the position of the saddle as given at the beginning of this section). At some point, the resonant periodic orbits are released to the unbound region. The dynamical influence of resonant periodic orbits depends strongly on their location and linear and nonlinear stability properties. Looking at Fig.~\ref{fig:StaticFieldApproximation} (inset), we note the elliptic island surrounding the $1$:$12$ resonant elliptic periodic orbit while the orbit is in the unbound region. Compared to Fig.~\ref{fig:PSOS_RESI1}, we note that this elliptic region corresponds to a part of phase space where no delayed ionization is detected. In the same figure, the resonant orbit~${\mathcal O}_{12}^h$ is hyperbolic. To analyze its nonlinear properties, we compute its stable and unstable manifolds and display the resulting pictures in Fig.~\ref{fig:PSOS_RESI1} (right panels). The manifolds are the pathways by which the electrons approach or leave the chaotic region. Note the strong similarities of the unstable manifold of $\mathcal{O}_{12}^h$ with the Poincar\'e sections of RESI trajectories in Fig.~\ref{fig:Real_RESI}. This similarity confirms the important role played by this unstable manifold (associated with a one-electron model) in the delayed double ionization process.

Looking at Fig.~\ref{fig:PSOS_RESI1}, we see that the stable and unstable manifolds associated with $\mathcal{O}_{n}^h$ develop in the vicinity of the bound region and branches extend deep into the unbound region. Additionally, the width of the chaotic region as given by those manifolds explains why the transition between the bound and unbound regions is so sharp~\cite{Maug09}. A comparison between the left and right panels of Fig.~\ref{fig:PSOS_RESI1} shows that the unstable manifold of $\mathcal{O}_{n}^h$ regulates the dynamics of the RESI trajectories. Even some of the very fine details of the unstable manifold are reproduced by this set of RESI trajectories. We note that some parts of the unstable manifold are missing: They correspond to the intersections between the stable and the unstable manifolds. The overlap between the stable and unstable manifolds of $\mathcal{O}_{n}^h$ forms a ``sticky'' region~\cite{OrderChaosDynAstro} that traps trajectories for some time before ionizing. 

\begin{figure}
	\centering
		\includegraphics[width = \linewidth]{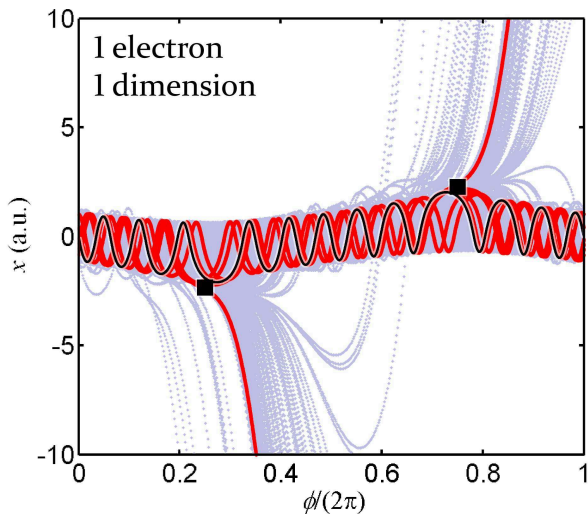}
	\caption{\label{fig:Manifold}
	(color online) Projection of the unstable manifold of the periodic orbit $\mathcal{O}_{12}^h$ of Hamiltonian~(\ref{eq:He+}) for $I=3\times 10^{15}\ {\rm W}~{\rm cm}^{-2}$ in the $(x,\phi)$ plane. Light gray corresponds to the manifold integrated for a long time while the dark gray focuses on the manifold integrated for a short time. The black curve is a projection of the periodic orbit $\mathcal{O}_{12}^h$ in the plane $(x,\phi)$, and the black squares indicate the position of the Stark saddle~\cite{ChaosAtomPhys} at the maximum of the field.}
\end{figure}

For deeper insight into the shape of the unstable manifold of $\mathcal{O}_n^h$, we represent a projection of this manifold in the plane $(x,\phi)$ together with a projection of $\mathcal{O}_n^h$ (black curve) in Fig.~\ref{fig:Manifold}. It shows a central region near $x=0$ and two main branches which depart from this central region. It is worth noticing that these branches are located near the extrema of the electric field, i.e., when $\phi=\pi/2$ and $\phi=3\pi/2$. Since a delayed ionization trajectory follows this unstable manifold asymptotically, we see that this trajectory leaves the nucleus at times that correspond to the extrema of the electric field. This is consistent with what is already observed for RESI~\cite{Ruiz08}. The unstable manifold has some other branches through which an electron can leave the nucleus and that do not correspond to extrema of the electric field. However, the probability of those ionization channels is much smaller. 

\begin{figure}
	\centering
		\includegraphics[width = \linewidth]{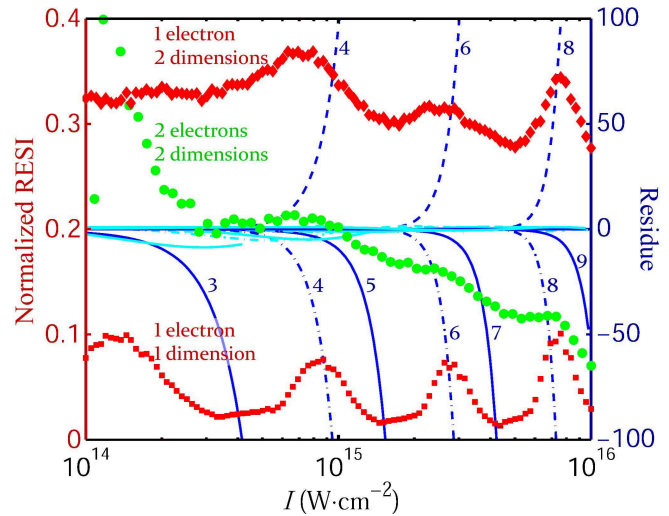}
	\caption{\label{fig:RESI_to_DI_ratio}
	(color online) Normalized RESI yields for 460~nm wavelength laser (markers, left hand axis). Square (diamond) markers label one (two) dimensional one-electron simulations (normalized to the number of nonionized trajectories). Dot markers label two-dimensional two-electron simulations (normalized to the number of double ionizations). The continuous and dashed curves represent the residue of the resonant periodic orbits $\mathcal{O}_{n}^{-}$ and $\mathcal{O}_{n}^{h}$ of Hamiltonian~(\ref{eq:He+}) (right hand axis) as a function of the laser intensity~$I$. Continuous (dashed) curves refer to odd (even) 1:$n$ resonances. As intensity increases, the order of the resonance goes from $n=3$ to $n=9$ as indicated on the curves.}
\end{figure}

The discussion of the laser intensity considered in Figs.~\ref{fig:PSOS_RESI1} and~\ref{fig:StaticFieldApproximation} can be generalized to arbitrary intensity. In this case, the hyperbolic resonant periodic orbit(s) in the unbound region (${\mathcal O}_{n}^{-}$ or ${\mathcal O}_{n}^h$ depending on the resonance index $n$) dynamically organize the chaotic layer surrounding the bound region, through their stable and unstable manifolds. The unbound region hosts many resonant orbits, associated with the accessible $m$:$n$ resonances. However, because of their short period (the same as the laser) and the short duration of the trapping for RESI (at most a couple of laser cycles) the primary resonances $1$:$n$ are the ones which organize the RESI dynamics. The other periodic orbits in the vicinity of the bound region influence only the fine details of the chaotic structure. As a signature of the influence of the $1$:$n$ resonances, we represent the normalized RESI statistics with respect to the double ionization yield as a function of the laser intensity for two-electron Hamiltonian~(\ref{eq:FullHamiltonian}) with $d=1$ (lower curve with square markers in Fig.~\ref{fig:RESI_to_DI_ratio}). We clearly see some oscillations as a function of the intensity. In order to correlate these oscillations with some features of the 1:$n$ resonances, we represent the value of the Greene's residue~\cite{Gree79} of $\mathcal{O}_{n}^{-}$ and $\mathcal{O}_{n}^{h}$ as a function of the intensity. These curves illustrate the strong correlation between the oscillations in the relative RESI yields with the successive releases of the resonant orbits in the unbound region. 

%%%%%%%%%% Dynamics of higher dimensional models %%%%%%%%%%%%%%%%%%%%%%%%%%%%%%%%%%%%%%%%%%%%%%%%%%%%%%%%%%%%%%%
\subsection{Dynamics of higher dimensional models} \label{sec:dim}

\begin{figure}
	\centering
		\includegraphics[width = \linewidth]{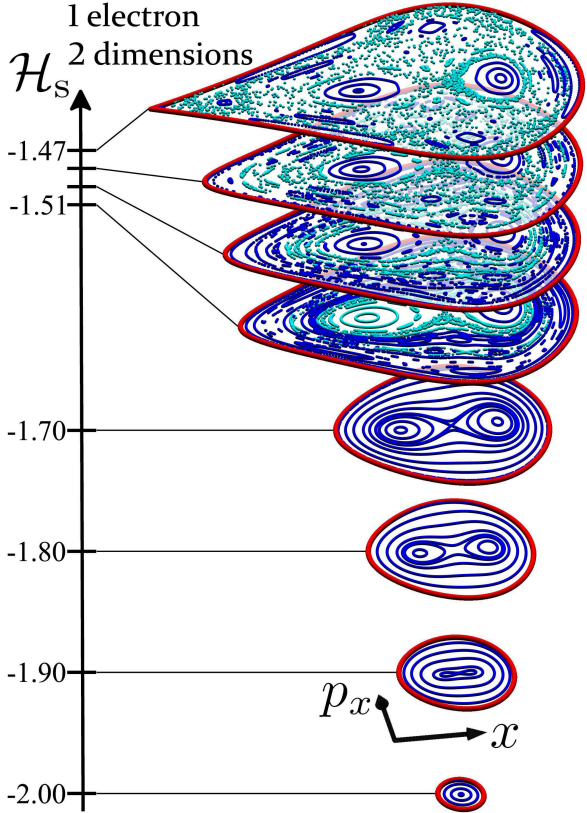}
	\caption{\label{fig:2dAdiabaticPS}
	(color online) Poincar\'e sections $y=0$ of Hamiltonian~(\ref{eq:HamAdia}) with $d=2$ for a laser intensity $I=3\times10^{15}\ {\rm W}~{\rm cm}^{-2}$ and $a=1$ and energy ranking from the Stark saddle value (top) to $\mathcal{H}_{s}=-2$. The outermost (red online) curve corresponds to the one dimensional model with the same parameters (see Fig.~\ref{fig:StaticFieldApproximation}).}
\end{figure}

\begin{figure}
	\centering
		\includegraphics[width = \linewidth]{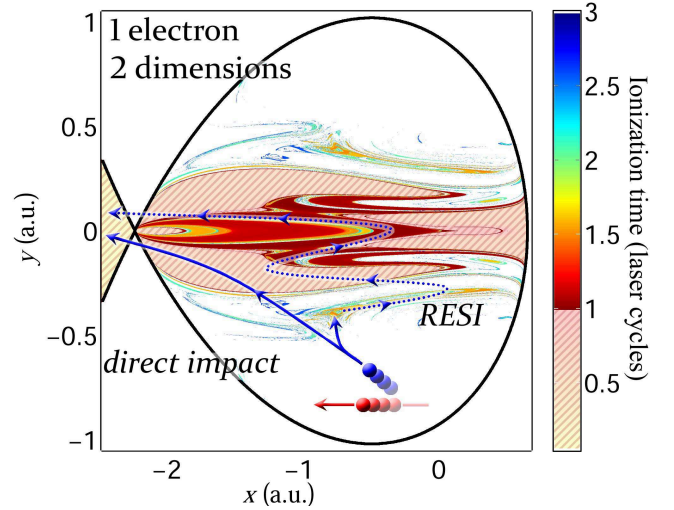}
	\caption{\label{fig:RESI_mechanism}
	(color online) Ionization time for Hamiltonian~(\ref{eq:He+}) with two dimensions for $I=3\times10^{15}\ {\rm W}~{\rm cm}^{-2}$, $\phi_{0}=\pi/2$ and $780$~nm. Initial conditions are chosen at the energy of the Stark saddle~\cite{ChaosAtomPhys} with $p_{x}\leq0$ and $p_{y}=0$. After a recollision (full line arrows), the inner electron can be directly ionized or promoted to an excited state that is ionized with a delay (dotted arrow).}
\end{figure}

In the adiabatic approximation, the one-electron model~(\ref{eq:HamAdia}) is integrable for $d=1$. A priori, this is not the case for two and three dimensions. In Fig.~\ref{fig:2dAdiabaticPS} we display the projection of the Poincar\'e section for Hamiltonian~(\ref{eq:HamAdia}) with $d=2$, corresponding to the adiabatic approximation of Hamiltonian~(\ref{eq:He+}). We see that the dynamics is a mixture of regular and chaotic trajectories, with a large chaotic sea [light gray (cyan online) dots on the Poincar\'e section] for energy values close to the Stark saddle. Since the laser is time independent, the separatrix (red curve) acts as a barrier to ionization, and trajectories are confined inside the bound region (on each horizontal plane). Compared to Hamiltonian~(\ref{eq:He+}) with $d=1$, the source of chaotic trajectories is very different: For Hamiltonian~(\ref{eq:HamAdia}) with $d=2$, it is the result of a nonlinear coupling between the two spatial dimensions, while for Hamiltonian~(\ref{eq:He+}) with $d=1$, it is the result of the interaction between the particle and the time-dependent field. It is expected that for the phase space of Hamiltonian~(\ref{eq:He+}) with $d=2$ these two sources of chaos both guide ionization. As a consequence, the dynamics allows for the possibility of an electron relatively deep inside the core region to diffuse chaotically and gain energy to be finally released after some delay. This phenomenon corresponds to the mechanism we have referred to as chaotic diffusion when no collision or soft interaction is involved in the subsequent ionization. Compared to Fig.~\ref{fig:2dAdiabaticPS}, the size of the chaotic region is amplified by the fact that additional tori are broken under the time dependence of the field. As a consequence, it becomes harder to define a bound and unbound region for the inner electron (see Fig.~\ref{fig:RESI_mechanism}), because of chaotic diffusion. 

As we did for the model~(\ref{eq:He+}) with $d=1$, the laser excitation can be considered as a perturbation of the free field dynamics for $d=2$. Then, again, KAM type arguments show the persistence of invariant tori in the vicinity of the core. However, contrary to the one dimensional case, these tori are dimensionally too small to create barriers of transport in phase space and a chaotic diffusion can happen across them. To aid this view, and in a first (rough) approximation, one can imagine all the light gray (cyan online) dot regions in Fig.~\ref{fig:2dAdiabaticPS} to be connected, such that by moving from one layer to the next it is possible to cruise from one chaotic region to another even though there is a torus in between the two. More realistically, the manifolds of the resonant orbits, through the transverse direction(s) to the the field, develop relatively deep inside the core region. This allows for chaotic diffusion, which is drastically reduced for one dimensional models due to the existence of the boundary torus. Close to the nucleus however, because KAM tori form an intricate network, and by continuity of the dynamics, the diffusion is extremely slow (as stated in Nekhoroshev's theorem~\cite{LectureNotesInPhysics}). As a result, and since we consider very short laser pulses, in practice it is possible to identify a (finite time) bound and unbound regions with a chaotic layer surrounding the bound region. Compared to the one dimensional case, this chaotic layer is more complicated, with branches developed deeper into the core region. 

In Fig.~\ref{fig:RESI_to_DI_ratio} we notice that the RESI yields given by Hamiltonian~(\ref{eq:FullHamiltonian}) are higher for $d=2$ than for $d=1$. We attribute these higher yields to the aforementioned chaotic dynamics coming from an increase of instability associated with additional resonances in higher dimensional models. It underscores once more the pivotal role played by the inner electron dynamics [Hamiltonian~(\ref{eq:He+})] in these delayed ionizations. 

\begin{figure}
	\centering
		\includegraphics[width = \linewidth]{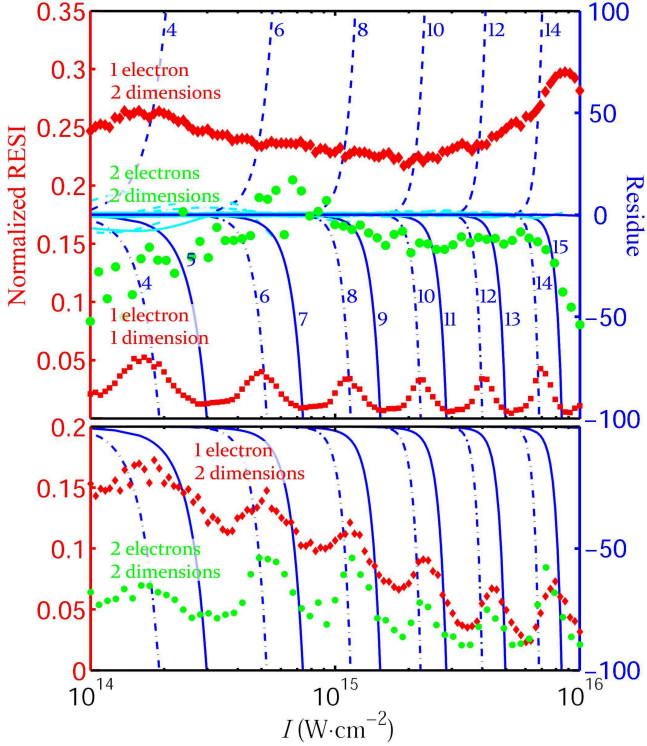}
	\caption{\label{fig:Oscillations780}
	(color online) Same figure as Fig.~\ref{fig:RESI_to_DI_ratio} for 780~nm wavelength laser. The upper panel corresponds to the standard soft Coulomb potential while the lower panel corresponds to the modified potential [see Hamiltonian~(\ref{eq:ModifiedFullHamiltonian})].}
\end{figure}

For two dimensional models a direct consequence of the unstable direction for the resonant periodic orbits in the direction transverse to the laser field is the smoothing of the oscillations in Fig.~\ref{fig:RESI_to_DI_ratio} for $460$~nm wavelength. Increasing the wavelength to $780$~nm gives birth to more resonances with a more dense tangle of unstable manifolds so that the oscillations are completely washed out in two-dimensional models whether they are one- or two-electron models, as observed from our classical calculations in Fig.~\ref{fig:Oscillations780} (upper panel).

A second consequence of the unstable direction in the transverse plane is the increase of delayed double ionization yields as can be seen in Fig.~\ref{fig:Oscillations780}. In the upper panel, the statistics for two dimensions (diamonds) is roughly one order of magnitude larger than it is for one dimension (squares). The same trend is confirmed with two electron models (compare Fig.~\ref{fig:Oscillations780}, dots in the upper panel, with Fig.~\ref{fig:RESI_to_DI_ratio_Atom}, filled in dots in the middle panel) with roughly one order of magnitude of difference between the one and two dimensional models. Looking at typical RESI trajectories we notice statistically longer delays for two dimensional models than for one. Again, we attribute these longer delays to the enhanced chaotic dynamics with two dimensions.

Going to higher dimensions the difference between two and three dimensional models is not as drastic as between one and two dimensions because the two dimensions perpendicular to the polarization direction are dynamically playing the same role due to the rotational symmetry of the system around the polarization direction.

To illustrate the importance on the dynamics played by the direction transverse to the field, we consider a two-dimensional model where the Coulomb potential is modified in the $y$-direction. We consider the following modification of Hamiltonian~(\ref{eq:FullHamiltonian})
\begin{widetext}
\begin{eqnarray}
  \tilde{\mathcal H} \left( {\bf x}_{1}, {\bf x}_{2}, {\bf p}_{1}, {\bf p}_{2}, t \right) & = &
      \frac{\left\|{\bf p}_{1}\right\|^{2}}{2} + \frac{\left\|{\bf p}_{2}\right\|^{2}}{2}
      - \frac{2}{\sqrt{\left\|\Omega{\bf x}_{1}\right\|^{2} + a^{2}}}
       - \frac{2}{\sqrt{\left\|\Omega{\bf x}_{2}\right\|^{2} + a^{2}}}
      + \frac{1}{\sqrt{\left\|{\bf x}_{1}-{\bf x}_{2}\right\|^{2} + b^{2}}}  \nonumber \\
     & & + \left( {\bf x}_{1} + {\bf x}_{2}\right) \cdot {\bf e}_{x} E_{0} f\left( t \right) \sin \omega t,  \label{eq:ModifiedFullHamiltonian}
\end{eqnarray}
\end{widetext}
where we $\Omega$ is the matrix
$$
   \Omega = \left(\begin{array}{cc} 1 & 0 \\ 0 & \alpha \end{array}\right).
$$
Tuning the parameter $\alpha$ enables one to modify the stability of the relevant resonant periodic orbit in the transverse direction without changing its location and stability in the polarization direction. For numerical simulations, we set $\alpha=2$ and display the result in Fig.~\ref{fig:Oscillations780} (lower panel). For such a parameter, the resonant periodic orbits become stable in the transverse direction. As a consequence of the stabilization of the transverse dynamics, the oscillations in the relative RESI yields are recovered (both with one and two electrons). Note also that the relative RESI yields are significantly reduced compared to Hamiltonian~(\ref{eq:FullHamiltonian}) which has a more chaotic dynamics.

%%%%%%%%%% Influence of the atom %%%%%%%%%%%%%%%%%%%%%%%%%%%%%%%%%%%%%%%%%%%%%%%%%%%%%%%%%%%%%%%%%%%%%%%%%%%%%%%
\subsection{Significance of the target atom} \label{sec:InfluenceOfTheAtom}

\begin{figure}
	\centering
		\includegraphics[width = \linewidth]{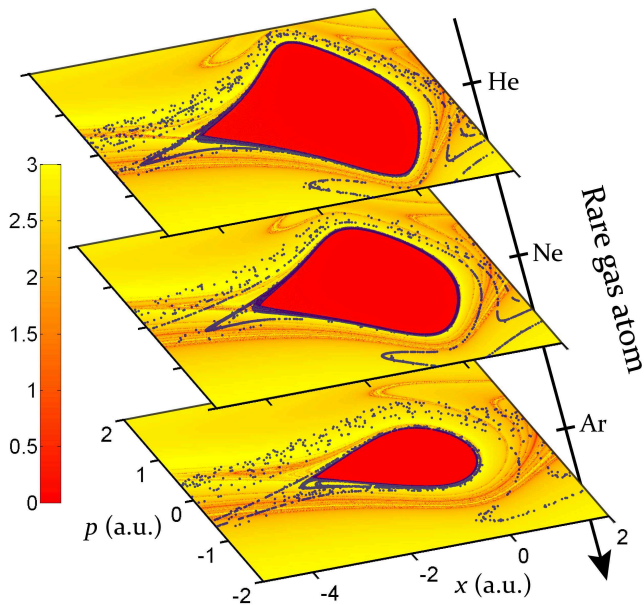}
	\caption{\label{fig:ComparePoincareSections}
	(color online) Poincar\'e section (dark dots, stroboscopic plots at the maximum of the laser field, i.e., $\phi=\pi/2$) of RESI trajectories of Hamiltonian~(\ref{eq:He+}) with $d=1$ for ${\rm He}^{+}$, ${\rm Ne}^{+}$ and ${\rm Ar}^{+}$ (from up to down) detected from a large set of random initial conditions for a laser intensity $I=10^{15} \ {\rm W}~{\rm cm}^{-2}$. On each panel, the color scale corresponds to laminar plots (distance, in logarithmic scale, of the electron from the core after $4.5$~laser cycles~\cite{Maug09}) for the same Hamiltonian and the considered atom.}
\end{figure}

When changing the target atom, there has been shown an apparent strong sensitivity of RESI yields to the considered atom~\cite{DeJe04_1, DeJe04_2, Rude04}. From the generic Hamiltonian~(\ref{eq:FullHamiltonian}) and its associated reduced models, these different atoms are modeled by varying the ground state energy and adjusting the softening parameters. Then, the analysis carried out in the previous sections can be applied to the specific atom. In particular, we focus on helium ($a=0.8$), neon ($a=1$) and argon ($a=1.5$) and use the same electron-electron softening parameter for the three of them ($b=1$). For instance, using Eq.~(\ref{eq:NbResonances}) we find $n_{\rm res}=33$ primary resonances for He, $n_{\rm res}=24$ for Ne and $n_{\rm res}=13$ for Ar so that the number of primary resonances $1$:$n$ decreases with $a$ as $a^{-3/2}$. For all atoms, we find a similar organization of the delayed ionization dynamics by the resonant orbits (see Fig.~\ref{fig:ComparePoincareSections}).

\begin{figure}
	\centering
		\includegraphics[width = \linewidth]{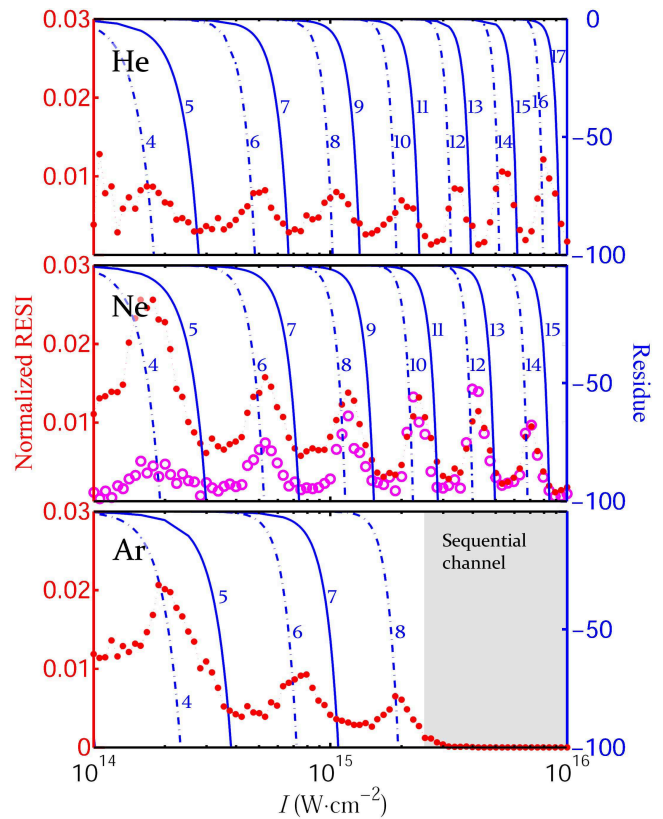}
	\caption{\label{fig:RESI_to_DI_ratio_Atom}
	(color online) Same figure as Fig.~\ref{fig:RESI_to_DI_ratio} for 780~nm wavelength laser for Hamiltonian~(\ref{eq:Ham2e}). In the panels we consider different atoms: He (upper, $a=0.8$), Ne (middle, $a=1$) and Ar (lower, $a=1.5$). In the middle panel we also compare the relative yield obtained with the non-aligned model~(\ref{eq:Eckh:Ham2e}) (open circles).}
\end{figure}

Previous studies have shown differences in the momentum spectra for various atoms~\cite{DeJe04_1, DeJe04_2, Rude04}. Our analysis suggests that this difference does not come from the phase space structures which are very similar for all atoms. To investigate the specific impact of the resonant orbits on delayed double ionization, we compare the RESI yields for each of the three atoms. The resulting picture is displayed in Fig.~\ref{fig:RESI_to_DI_ratio_Atom} for Hamiltonian~(\ref{eq:FullHamiltonian}) and $d=1$. As observed previously, in the nonsequential double ionization regime, the proportion of RESI exhibits oscillations. These oscillations are a clear signature of the resonant periodic orbits of the reduced Hamiltonian~(\ref{eq:He+}) which organize the chaotic layer around the bound region. Note the very good correlation between these oscillations and the residue curve of the $1$:$n$ resonant orbits for the three atoms. To investigate the importance of the aligned versus non-aligned model in delayed double ionization, in the middle panel we compare the RESI yields for Hamiltonians~(\ref{eq:Ham2e}) and~(\ref{eq:Eckh:Ham2e}). As expected, since they share the same inner electron reduced model, we observe the same trends between the two models where only the amplitude for low intensities is different. We also note that while similar oscillations are observed for the three atoms, the relative RESI yield is larger for Ne than it is it for both He and Ar, in agreement with what has been observed experimentally~\cite{DeJe04_2}.

%%%%%%%%%%%%%%%%%%%%%%%%%%%%%%%%%%%%%%%%%%%%%%%%%%%%%%%%%%%%%%%%%%%%%%%%%%%%%%%%%%%%%%%%%%%%%%%%%%%%%%%%%%%%%%%%
%%                                                 Conclusion                                                 %%
%%%%%%%%%%%%%%%%%%%%%%%%%%%%%%%%%%%%%%%%%%%%%%%%%%%%%%%%%%%%%%%%%%%%%%%%%%%%%%%%%%%%%%%%%%%%%%%%%%%%%%%%%%%%%%%%
\section{Conclusion}

We have shown that delayed double ionization results from Hamiltonian chaos. This chaotic behavior results from well-identified resonances between the free field dynamics and the laser excitation. Depending on the specific question at hand, we have shown the limits of the reduced models:
\begin{itemize}
   \item When the preionized electron is far away from the nucleus, the dynamics of the remaining electron is accurately described by the inner electron reduced model (from which the resonances are identified). As a consequence, different two electron reduced models that share the same inner electron model [see Hamiltonians~(\ref{eq:Ham2e}) and~(\ref{eq:Eckh:Ham2e})] lead to qualitatively similar results for delayed double ionization.
   \item While one-dimensional reduced models have been successful in describing direct impact double ionization, they give a partial description for the mechanisms behind delayed double ionization. In this case, they exhibit the skeleton that regulates delayed ionization through the 1:$n$ resonances. However, they miss the influence of the transverse dimensions that enhance chaos. As a consequence, one dimensional models tend to significantly underestimate both the amount and the delays for delayed double ionization.
   \item The mechanism of resonance is very robust against changes of parameters (laser frequency and species). In particular, for different atoms, the number of resonances varies, leading to a slight variation in normalized RESI yields. Nevertheless, in all cases, the resonances are correlated in the same way to the observed oscillations in the relative RESI yields.
\end{itemize}
All of these limits have been illustrated, e.g., on the plots of the normalized RESI yields versus intensity. In particular, in some range of parameters, it has been shown that these curves exhibit oscillations which act as a clear signature of Hamiltonian chaos in these systems.

\begin{acknowledgments}
The authors acknowledge useful discussions with C. Cirelli, R.~Doerner, N.~Johnson, R.~R.~Jones, U.~Keller, M.~Kling, R.~Moshammer, and A.~Pfeiffer. C.C. and F.M. acknowledge financial support from the CNRS. F.M. acknowledges financial support from the Fulbright program. This work is partially funded by NSF.
\end{acknowledgments}

%%%%%%%%%%%%%%%%%%%%%%%%%%%%%%%%%%%%%%%%%%%%%%%%%%%%%%%%%%%%%%%%%%%%%%%%%%%%%%%%%%%%%%%%%%%%%%%%%%%%%%%%%%%%%%%%
%%                                                BIBLIOGRAPHY                                                %%
%%%%%%%%%%%%%%%%%%%%%%%%%%%%%%%%%%%%%%%%%%%%%%%%%%%%%%%%%%%%%%%%%%%%%%%%%%%%%%%%%%%%%%%%%%%%%%%%%%%%%%%%%%%%%%%%
%\section*{References}

% Bibliography
%\bibliographystyle{unsrt}
%\bibliographystyle{apsrev}
%\bibliography{bibliography} 

%merlin.mbs 2010-03-15 4.21a (PWD, AO, DPC)
\begin{thebibliography}{0}%
\makeatletter
\providecommand \@ifxundefined [1]{%
 \@ifx{#1\undefined}
}%
\providecommand \@ifnum [1]{%
 \ifnum #1\expandafter \@firstoftwo
 \else \expandafter \@secondoftwo
 \fi
}%
\providecommand \@ifx [1]{%
 \ifx #1\expandafter \@firstoftwo
 \else \expandafter \@secondoftwo
 \fi
}%
\providecommand \natexlab [1]{#1}%
\providecommand \enquote  [1]{``#1''}%
\providecommand \bibnamefont  [1]{#1}%
\providecommand \bibfnamefont [1]{#1}%
\providecommand \citenamefont [1]{#1}%
\providecommand \href@noop [0]{\@secondoftwo}%
\providecommand \href [0]{\begingroup \@sanitize@url \@href}%
\providecommand \@href[1]{\@@startlink{#1}\@@href}%
\providecommand \@@href[1]{\endgroup#1\@@endlink}%
\providecommand \@sanitize@url [0]{\catcode `\\12\catcode `\$12\catcode
  `\&12\catcode `\#12\catcode `\^12\catcode `\_12\catcode `\%12\relax}%
\providecommand \@@startlink[1]{}%
\providecommand \@@endlink[0]{}%
\providecommand \url  [0]{\begingroup\@sanitize@url \@url }%
\providecommand \@url [1]{\endgroup\@href {#1}{\urlprefix }}%
\providecommand \urlprefix  [0]{URL }%
\providecommand \Eprint [0]{\href }%
\@ifxundefined \urlstyle {%
  \providecommand \doi  [0]{\begingroup \@sanitize@url \@doi}%
  \providecommand \@doi [1]{\endgroup \@@startlink {\doibase
  #1}doi:\discretionary {}{}{}#1\@@endlink }%
}{%
  \providecommand \doi  [0]{doi:\discretionary{}{}{}\begingroup
  \urlstyle{rm}\Url }%
}%
\providecommand \doibase [0]{http://dx.doi.org/}%
\providecommand \Doi [0]{\begingroup \@sanitize@url \@Doi }%
\providecommand \@Doi  [1]{\endgroup\@@startlink{\doibase#1}\@@Doi}%
\providecommand \@@Doi [1]{#1\@@endlink}%
\providecommand \selectlanguage [0]{\@gobble}%
\providecommand \bibinfo  [0]{\@secondoftwo}%
\providecommand \bibfield  [0]{\@secondoftwo}%
\providecommand \translation [1]{[#1]}%
\providecommand \BibitemOpen [0]{}%
\providecommand \bibitemStop [0]{}%
\providecommand \bibitemNoStop [0]{.\EOS\space}%
\providecommand \EOS [0]{\spacefactor3000\relax}%
\providecommand \BibitemShut  [1]{\csname bibitem#1\endcsname}%
%</preamble>
\end{thebibliography}%


\begin{thebibliography}{33}
\expandafter\ifx\csname natexlab\endcsname\relax\def\natexlab#1{#1}\fi
\expandafter\ifx\csname bibnamefont\endcsname\relax
  \def\bibnamefont#1{#1}\fi
\expandafter\ifx\csname bibfnamefont\endcsname\relax
  \def\bibfnamefont#1{#1}\fi
\expandafter\ifx\csname citenamefont\endcsname\relax
  \def\citenamefont#1{#1}\fi
\expandafter\ifx\csname url\endcsname\relax
  \def\url#1{\texttt{#1}}\fi
\expandafter\ifx\csname urlprefix\endcsname\relax\def\urlprefix{URL }\fi
\providecommand{\bibinfo}[2]{#2}
\providecommand{\eprint}[2][]{\url{#2}}

\bibitem[{\citenamefont{Becker and Rottke}(2008)}]{Beck08}
\bibinfo{author}{\bibfnamefont{W.}~\bibnamefont{Becker}} \bibnamefont{and}
  \bibinfo{author}{\bibfnamefont{H.}~\bibnamefont{Rottke}},
  \bibinfo{journal}{Contemporary Physics} \textbf{\bibinfo{volume}{49}},
  \bibinfo{pages}{199} (\bibinfo{year}{2008}).

\bibitem[{\citenamefont{Corkum}(1993)}]{Cork93}
\bibinfo{author}{\bibfnamefont{P.~B.} \bibnamefont{Corkum}},
  \bibinfo{journal}{Phys.~Rev.~Lett.} \textbf{\bibinfo{volume}{71}},
  \bibinfo{pages}{1994} (\bibinfo{year}{1993}).

\bibitem[{\citenamefont{Schafer et~al.}(1993)\citenamefont{Schafer, Yang,
  DiMauro, and Kulander}}]{Scha93}
\bibinfo{author}{\bibfnamefont{K.~J.} \bibnamefont{Schafer}},
  \bibinfo{author}{\bibfnamefont{B.}~\bibnamefont{Yang}},
  \bibinfo{author}{\bibfnamefont{L.~F.} \bibnamefont{DiMauro}},
  \bibnamefont{and} \bibinfo{author}{\bibfnamefont{K.~C.}
  \bibnamefont{Kulander}}, \bibinfo{journal}{Phys.~Rev.~Lett.}
  \textbf{\bibinfo{volume}{70}}, \bibinfo{pages}{1599} (\bibinfo{year}{1993}).

\bibitem[{\citenamefont{Rudenko et~al.}(2004)\citenamefont{Rudenko, Zrost,
  Feuerstein, de~Jesus, Schr\"oter, Moshammer, and Ullrich}}]{Rude04}
\bibinfo{author}{\bibfnamefont{A.}~\bibnamefont{Rudenko}},
  \bibinfo{author}{\bibfnamefont{K.}~\bibnamefont{Zrost}},
  \bibinfo{author}{\bibfnamefont{B.}~\bibnamefont{Feuerstein}},
  \bibinfo{author}{\bibfnamefont{V.~L.~B.} \bibnamefont{de~Jesus}},
  \bibinfo{author}{\bibfnamefont{C.~D.} \bibnamefont{Schr\"oter}},
  \bibinfo{author}{\bibfnamefont{R.}~\bibnamefont{Moshammer}},
  \bibnamefont{and} \bibinfo{author}{\bibfnamefont{J.}~\bibnamefont{Ullrich}},
  \bibinfo{journal}{Phys.~Rev.~Lett.} \textbf{\bibinfo{volume}{93}},
  \bibinfo{pages}{253001} (\bibinfo{year}{2004}).

\bibitem[{\citenamefont{Feuerstein et~al.}(2001)\citenamefont{Feuerstein,
  Moshammer, Fischer, Dorn, Schr\"oter, Deipenwisch, Crespo Lopez-Urrutia,
  H\"ohr, Neumayer, Ullrich, Rottke, Trump, Wittmann, Korn, and Sandner}}]{Feue01}
\bibinfo{author}{\bibfnamefont{B.}~\bibnamefont{Feuerstein}},
  \bibinfo{author}{\bibfnamefont{R.}~\bibnamefont{Moshammer}},
  \bibinfo{author}{\bibfnamefont{D.}~\bibnamefont{Fischer}},
  \bibinfo{author}{\bibfnamefont{A.}~\bibnamefont{Dorn}},
  \bibinfo{author}{\bibfnamefont{C.~D.} \bibnamefont{Schr\"oter}},
  \bibinfo{author}{\bibfnamefont{J.}~\bibnamefont{Deipenwisch}},
  \bibinfo{author}{\bibfnamefont{J.~R.} \bibnamefont{Crespo Lopez-Urrutia}},
  \bibinfo{author}{\bibfnamefont{C.}~\bibnamefont{H\"ohr}},
  \bibinfo{author}{\bibfnamefont{P.}~\bibnamefont{Neumayer}},
  \bibinfo{author}{\bibfnamefont{J.}~\bibnamefont{Ullrich}},
  \bibinfo{author}{\bibfnamefont{H.}~\bibnamefont{Rottke}},
  \bibinfo{author}{\bibfnamefont{C.}~\bibnamefont{Trump}},
  \bibinfo{author}{\bibfnamefont{M.}~\bibnamefont{Wittmann}},
  \bibinfo{author}{\bibfnamefont{G.}~\bibnamefont{Korn}},
  \bibinfo{author}{\bibfnamefont{W.}~\bibnamefont{Sandner}},
  \bibinfo{journal}{Phys.~Rev.~Lett.}
  \textbf{\bibinfo{volume}{87}}, \bibinfo{pages}{043003}
  (\bibinfo{year}{2001}).

\bibitem[{\citenamefont{de~Jesus
  et~al.}(2004{\natexlab{a}})\citenamefont{de~Jesus, Rudenko, Feuerstein,
  Zrost, Schr\"{o}ter, Moshammer, and Ullrich}}]{DeJe04_1}
\bibinfo{author}{\bibfnamefont{V.~L.~B.} \bibnamefont{de~Jesus}},
  \bibinfo{author}{\bibfnamefont{A.}~\bibnamefont{Rudenko}},
  \bibinfo{author}{\bibfnamefont{B.}~\bibnamefont{Feuerstein}},
  \bibinfo{author}{\bibfnamefont{K.}~\bibnamefont{Zrost}},
  \bibinfo{author}{\bibfnamefont{C.~D.} \bibnamefont{Schr\"{o}ter}},
  \bibinfo{author}{\bibfnamefont{R.}~\bibnamefont{Moshammer}},
  \bibnamefont{and} \bibinfo{author}{\bibfnamefont{J.}~\bibnamefont{Ullrich}},
  \bibinfo{journal}{J.~Elec.~Spect.} \textbf{\bibinfo{volume}{141}},
  \bibinfo{pages}{127} (\bibinfo{year}{2004}{\natexlab{a}}).

\bibitem[{\citenamefont{de~Jesus
  et~al.}(2004{\natexlab{b}})\citenamefont{de~Jesus, Feuerstein, Zrost,
  Fischer, Rudenko, Afaneh, Schr\"{o}ter, Moshammer, and Ullrich}}]{DeJe04_2}
\bibinfo{author}{\bibfnamefont{V.~L.~B.} \bibnamefont{de~Jesus}},
  \bibinfo{author}{\bibfnamefont{B.}~\bibnamefont{Feuerstein}},
  \bibinfo{author}{\bibfnamefont{K.}~\bibnamefont{Zrost}},
  \bibinfo{author}{\bibfnamefont{D.}~\bibnamefont{Fischer}},
  \bibinfo{author}{\bibfnamefont{A.}~\bibnamefont{Rudenko}},
  \bibinfo{author}{\bibfnamefont{F.}~\bibnamefont{Afaneh}},
  \bibinfo{author}{\bibfnamefont{C.~D.} \bibnamefont{Schr\"{o}ter}},
  \bibinfo{author}{\bibfnamefont{R.}~\bibnamefont{Moshammer}},
  \bibnamefont{and} \bibinfo{author}{\bibfnamefont{J.}~\bibnamefont{Ullrich}},
  \bibinfo{journal}{J.~Phys.~B.} \textbf{\bibinfo{volume}{37}},
  \bibinfo{pages}{L161} (\bibinfo{year}{2004}{\natexlab{b}}).

\bibitem[{\citenamefont{van~de Sand and Rost}(1999)}]{Sand99}
\bibinfo{author}{\bibfnamefont{G.}~\bibnamefont{van~de Sand}} \bibnamefont{and}
  \bibinfo{author}{\bibfnamefont{J.~M.} \bibnamefont{Rost}},
  \bibinfo{journal}{Phys.~Rev.~Lett.} \textbf{\bibinfo{volume}{83}},
  \bibinfo{pages}{524} (\bibinfo{year}{1999}).

\bibitem[{\citenamefont{Figueira~de Morisson~Faria and Liu}(2011)}]{Figu11}
\bibinfo{author}{\bibfnamefont{C.}~\bibnamefont{Figueira~de Morisson~Faria}}
  \bibnamefont{and} \bibinfo{author}{\bibfnamefont{X.}~\bibnamefont{Liu}},
  \bibinfo{journal}{J.~Modern.~Opt.} \textbf{\bibinfo{volume}{58}},
  \bibinfo{pages}{1076} (\bibinfo{year}{2011}).

\bibitem[{\citenamefont{Weber et~al.}(2000)\citenamefont{Weber, Weckenbrock,
  Staudte, Spielberger, Jagutzki, Mergel, Afaneh, Urbasch, Vollmer, Giessen,
  and D\"orner}}]{Webe00_2}
\bibinfo{author}{\bibfnamefont{T.}~\bibnamefont{Weber}},
  \bibinfo{author}{\bibfnamefont{M.}~\bibnamefont{Weckenbrock}},
  \bibinfo{author}{\bibfnamefont{A.}~\bibnamefont{Staudte}},
  \bibinfo{author}{\bibfnamefont{L.}~\bibnamefont{Spielberger}},
  \bibinfo{author}{\bibfnamefont{O.}~\bibnamefont{Jagutzki}},
  \bibinfo{author}{\bibfnamefont{V.}~\bibnamefont{Mergel}},
  \bibinfo{author}{\bibfnamefont{F.}~\bibnamefont{Afaneh}},
  \bibinfo{author}{\bibfnamefont{G.}~\bibnamefont{Urbasch}},
  \bibinfo{author}{\bibfnamefont{M.}~\bibnamefont{Vollmer}},
  \bibinfo{author}{\bibfnamefont{H.}~\bibnamefont{Giessen}},
  \bibinfo{author}{\bibfnamefont{R.}~\bibnamefont{D\"orner}},
  \bibinfo{journal}{Phys.~Rev.~Lett.}
  \textbf{\bibinfo{volume}{84}}, \bibinfo{pages}{443} (\bibinfo{year}{2000}).

\bibitem[{\citenamefont{Moshammer et~al.}(2000)\citenamefont{Moshammer,
  Feuerstein, Schmitt, Dorn, Schr\"oter, Ullrich, Rottke, Trump, Wittmann, Korn,
  Hoffmann, and Sandner}}]{Mosh00}
\bibinfo{author}{\bibfnamefont{R.}~\bibnamefont{Moshammer}},
  \bibinfo{author}{\bibfnamefont{B.}~\bibnamefont{Feuerstein}},
  \bibinfo{author}{\bibfnamefont{W.}~\bibnamefont{Schmitt}},
  \bibinfo{author}{\bibfnamefont{A.}~\bibnamefont{Dorn}},
  \bibinfo{author}{\bibfnamefont{C.~D.} \bibnamefont{Schr\"oter}},
  \bibinfo{author}{\bibfnamefont{J.}~\bibnamefont{Ullrich}},
  \bibinfo{author}{\bibfnamefont{H.}~\bibnamefont{Rottke}},
  \bibinfo{author}{\bibfnamefont{C.}~\bibnamefont{Trump}},
  \bibinfo{author}{\bibfnamefont{M.}~\bibnamefont{Wittmann}},
  \bibinfo{author}{\bibfnamefont{G.}~\bibnamefont{Korn}},
  \bibinfo{author}{\bibfnamefont{K.}~\bibnamefont{Hoffmann}},
  \bibinfo{author}{\bibfnamefont{W.}~\bibnamefont{Sandner}},
  \bibinfo{journal}{Phys.~Rev.~Lett.} \textbf{\bibinfo{volume}{84}},
  \bibinfo{pages}{447} (\bibinfo{year}{2000}).

\bibitem[{\citenamefont{Ruiz and Becker}(2008)}]{Ruiz08}
\bibinfo{author}{\bibfnamefont{C.}~\bibnamefont{Ruiz}} \bibnamefont{and}
  \bibinfo{author}{\bibfnamefont{A.}~\bibnamefont{Becker}},
  \bibinfo{journal}{New~J.~Phys.} \textbf{\bibinfo{volume}{10}},
  \bibinfo{pages}{025020} (\bibinfo{year}{2008}).

\bibitem[{\citenamefont{Ivanov et~al.}(2005)\citenamefont{Ivanov, Spanner, and
  Smirnova}}]{Ivan05}
\bibinfo{author}{\bibfnamefont{M.~Y.} \bibnamefont{Ivanov}},
  \bibinfo{author}{\bibfnamefont{M.}~\bibnamefont{Spanner}}, \bibnamefont{and}
  \bibinfo{author}{\bibfnamefont{O.}~\bibnamefont{Smirnova}},
  \bibinfo{journal}{J.~Mod.~Opt.} \textbf{\bibinfo{volume}{52}},
  \bibinfo{pages}{165} (\bibinfo{year}{2005}).

\bibitem[{\citenamefont{Haan et~al.}(2008)\citenamefont{Haan, Van~Dyke, and
  Smith}}]{Haan08}
\bibinfo{author}{\bibfnamefont{S.~L.} \bibnamefont{Haan}},
  \bibinfo{author}{\bibfnamefont{J.~S.} \bibnamefont{Van~Dyke}},
  \bibnamefont{and} \bibinfo{author}{\bibfnamefont{Z.~S.} \bibnamefont{Smith}},
  \bibinfo{journal}{Phys.~Rev.~Lett.} \textbf{\bibinfo{volume}{101}},
  \bibinfo{pages}{113001} (\bibinfo{year}{2008}).

\bibitem[{\citenamefont{Shomsky et~al.}(2009)\citenamefont{Shomsky, Smith, and
  Haan}}]{Shom09}
\bibinfo{author}{\bibfnamefont{K.~N.} \bibnamefont{Shomsky}},
  \bibinfo{author}{\bibfnamefont{Z.~S.} \bibnamefont{Smith}}, \bibnamefont{and}
  \bibinfo{author}{\bibfnamefont{S.~L.} \bibnamefont{Haan}},
  \bibinfo{journal}{Phys.~Rev.~A} \textbf{\bibinfo{volume}{79}},
  \bibinfo{pages}{061402(R)} (\bibinfo{year}{2009}).

\bibitem[{\citenamefont{Haan et~al.}(2010)\citenamefont{Haan, Smith, Shomsky,
  Plantinga, and Atallah}}]{Haan10}
\bibinfo{author}{\bibfnamefont{S.~L.} \bibnamefont{Haan}},
  \bibinfo{author}{\bibfnamefont{Z.~S.} \bibnamefont{Smith}},
  \bibinfo{author}{\bibfnamefont{K.~N.} \bibnamefont{Shomsky}},
  \bibinfo{author}{\bibfnamefont{P.~W.} \bibnamefont{Plantinga}},
  \bibnamefont{and} \bibinfo{author}{\bibfnamefont{T.~L.}
  \bibnamefont{Atallah}}, \bibinfo{journal}{Phys.~Rev.~A}
  \textbf{\bibinfo{volume}{81}}, \bibinfo{pages}{023409}
  (\bibinfo{year}{2010}).

\bibitem[{\citenamefont{Mauger et~al.}(2012)\citenamefont{Mauger, Kamor,
  Chandre, and Uzer}}]{Maug12_1}
\bibinfo{author}{\bibfnamefont{F.}~\bibnamefont{Mauger}},
  \bibinfo{author}{\bibfnamefont{A.}~\bibnamefont{Kamor}},
  \bibinfo{author}{\bibfnamefont{C.}~\bibnamefont{Chandre}}, \bibnamefont{and}
  \bibinfo{author}{\bibfnamefont{T.}~\bibnamefont{Uzer}},
  \bibinfo{journal}{Phys.~Rev.~Lett.}
  \textbf{\bibinfo{volume}{108}}, \bibinfo{pages}{063001}
  (\bibinfo{year}{2012}).

\bibitem[{\citenamefont{Pfeiffer et~al.}(2011)\citenamefont{Pfeiffer, Cirelli,
  Smolarski, Wang, Eberly, D\"{o}rner, and Keller}}]{Pfei11}
\bibinfo{author}{\bibfnamefont{A.~N.} \bibnamefont{Pfeiffer}},
  \bibinfo{author}{\bibfnamefont{C.}~\bibnamefont{Cirelli}},
  \bibinfo{author}{\bibfnamefont{M.}~\bibnamefont{Smolarski}},
  \bibinfo{author}{\bibfnamefont{X.}~\bibnamefont{Wang}},
  \bibinfo{author}{\bibfnamefont{J.~H.} \bibnamefont{Eberly}},
  \bibinfo{author}{\bibfnamefont{R.}~\bibnamefont{D\"{o}rner}}, \bibnamefont{and}
  \bibinfo{author}{\bibfnamefont{U.}~\bibnamefont{Keller}},
  \bibinfo{journal}{New~J.~Phys.} \textbf{\bibinfo{volume}{13}},
  \bibinfo{pages}{093008} (\bibinfo{year}{2011}).

\bibitem[{\citenamefont{Panfili et~al.}(2001)\citenamefont{Panfili, Eberly, and
  Haan}}]{Panf01}
\bibinfo{author}{\bibfnamefont{R.}~\bibnamefont{Panfili}},
  \bibinfo{author}{\bibfnamefont{J.~H.} \bibnamefont{Eberly}},
  \bibnamefont{and} \bibinfo{author}{\bibfnamefont{S.~L.} \bibnamefont{Haan}},
  \bibinfo{journal}{Opt.~Express} \textbf{\bibinfo{volume}{8}},
  \bibinfo{pages}{431} (\bibinfo{year}{2001}).

\bibitem[{\citenamefont{Ho et~al.}(2005)\citenamefont{Ho, Panfili, Haan, and
  Eberly}}]{Ho05_1}
\bibinfo{author}{\bibfnamefont{P.~J.} \bibnamefont{Ho}},
  \bibinfo{author}{\bibfnamefont{R.}~\bibnamefont{Panfili}},
  \bibinfo{author}{\bibfnamefont{S.~L.} \bibnamefont{Haan}}, \bibnamefont{and}
  \bibinfo{author}{\bibfnamefont{J.~H.} \bibnamefont{Eberly}},
  \bibinfo{journal}{Phys.~Rev.~Lett.} \textbf{\bibinfo{volume}{94}},
  \bibinfo{pages}{093002} (\bibinfo{year}{2005}).

\bibitem[{\citenamefont{Mauger et~al.}(2009{\natexlab{a}})\citenamefont{Mauger,
  Chandre, and Uzer}}]{Maug09}
\bibinfo{author}{\bibfnamefont{F.}~\bibnamefont{Mauger}},
  \bibinfo{author}{\bibfnamefont{C.}~\bibnamefont{Chandre}}, \bibnamefont{and}
  \bibinfo{author}{\bibfnamefont{T.}~\bibnamefont{Uzer}},
  \bibinfo{journal}{Phys.~Rev.~Lett.} \textbf{\bibinfo{volume}{102}},
  \bibinfo{pages}{173002} (\bibinfo{year}{2009}{\natexlab{a}});
  \bibinfo{journal}{J.~Phys.~B.} \textbf{\bibinfo{volume}{42}},
  \bibinfo{pages}{165602} (\bibinfo{year}{2009}{\natexlab{b}}).

\bibitem[{\citenamefont{Mauger et~al.}(2010{\natexlab{a}})\citenamefont{Mauger,
  Chandre, and Uzer}}]{Maug10}
\bibinfo{author}{\bibfnamefont{F.}~\bibnamefont{Mauger}},
  \bibinfo{author}{\bibfnamefont{C.}~\bibnamefont{Chandre}}, \bibnamefont{and}
  \bibinfo{author}{\bibfnamefont{T.}~\bibnamefont{Uzer}},
  \bibinfo{journal}{Phys.~Rev.~Lett.} \textbf{\bibinfo{volume}{104}},
  \bibinfo{pages}{043005} (\bibinfo{year}{2010}{\natexlab{a}});
  \bibinfo{journal}{Phys.~Rev.~A} \textbf{\bibinfo{volume}{81}},
  \bibinfo{pages}{063425} (\bibinfo{year}{2010}{\natexlab{b}}).

\bibitem[{\citenamefont{Javanainen et~al.}(1988)\citenamefont{Javanainen,
  Eberly, and Su}}]{Java88}
\bibinfo{author}{\bibfnamefont{J.}~\bibnamefont{Javanainen}},
  \bibinfo{author}{\bibfnamefont{J.~H.} \bibnamefont{Eberly}},
  \bibnamefont{and} \bibinfo{author}{\bibfnamefont{Q.}~\bibnamefont{Su}},
  \bibinfo{journal}{Phys.~Rev.~A} \textbf{\bibinfo{volume}{38}},
  \bibinfo{pages}{3430} (\bibinfo{year}{1988}).

\bibitem[{\citenamefont{Panfili et~al.}(2002)\citenamefont{Panfili, Haan, and
  Eberly}}]{Panf02}
\bibinfo{author}{\bibfnamefont{R.}~\bibnamefont{Panfili}},
  \bibinfo{author}{\bibfnamefont{S.~L.} \bibnamefont{Haan}}, \bibnamefont{and}
  \bibinfo{author}{\bibfnamefont{J.~H.} \bibnamefont{Eberly}},
  \bibinfo{journal}{Phys.~Rev.~Lett.} \textbf{\bibinfo{volume}{89}},
  \bibinfo{pages}{113001} (\bibinfo{year}{2002}).

\bibitem[{\citenamefont{Haan et~al.}(1994)\citenamefont{Haan, Grobe, and
  Eberly}}]{Haan94}
\bibinfo{author}{\bibfnamefont{S.~L.} \bibnamefont{Haan}},
  \bibinfo{author}{\bibfnamefont{R.}~\bibnamefont{Grobe}}, \bibnamefont{and}
  \bibinfo{author}{\bibfnamefont{J.~H.} \bibnamefont{Eberly}},
  \bibinfo{journal}{Phys.~Rev.~A} \textbf{\bibinfo{volume}{50}},
  \bibinfo{pages}{378} (\bibinfo{year}{1994}).

\bibitem[{\citenamefont{Dirac}(1950)}]{Dira50}
\bibinfo{author}{\bibfnamefont{P.~A.~M.} \bibnamefont{Dirac}},
  \bibinfo{journal}{Can.~J.~Math.} \textbf{\bibinfo{volume}{2}},
  \bibinfo{pages}{129} (\bibinfo{year}{1950}).

\bibitem[{\citenamefont{Prauzner-Bechcicki
  et~al.}(2005)\citenamefont{Prauzner-Bechcicki, Sacha, Eckhardt, and
  Zakrzewski}}]{Prau05}
\bibinfo{author}{\bibfnamefont{J.~S.} \bibnamefont{Prauzner-Bechcicki}},
  \bibinfo{author}{\bibfnamefont{K.}~\bibnamefont{Sacha}},
  \bibinfo{author}{\bibfnamefont{B.}~\bibnamefont{Eckhardt}}, \bibnamefont{and}
  \bibinfo{author}{\bibfnamefont{J.}~\bibnamefont{Zakrzewski}},
  \bibinfo{journal}{Phys.~Rev.~A} \textbf{\bibinfo{volume}{71}},
  \bibinfo{pages}{033407} (\bibinfo{year}{2005}).

\bibitem[{\citenamefont{Eckhardt and Sacha}(2006)}]{Eckh06}
\bibinfo{author}{\bibfnamefont{B.}~\bibnamefont{Eckhardt}} \bibnamefont{and}
  \bibinfo{author}{\bibfnamefont{K.}~\bibnamefont{Sacha}},
  \bibinfo{journal}{J.~Phys.~B.} \textbf{\bibinfo{volume}{39}},
  \bibinfo{pages}{3865} (\bibinfo{year}{2006}).

\bibitem[{\citenamefont{Bl\"{u}mel and Reinhardt}(1997)}]{ChaosAtomPhys}
\bibinfo{author}{\bibfnamefont{R.}~\bibnamefont{Bl\"{u}mel}} \bibnamefont{and}
  \bibinfo{author}{\bibfnamefont{W.~P.} \bibnamefont{Reinhardt}},
  \emph{\bibinfo{title}{Chaos in Atomic Physics}}
  (\bibinfo{publisher}{Cambridge University Press, Cambridge, U.K.},
  \bibinfo{year}{1997}).

\bibitem[{\citenamefont{Lichtenberg and
  Lieberman}(1982)}]{RegAndStochastMotion}
\bibinfo{author}{\bibfnamefont{A.}~\bibnamefont{Lichtenberg}} \bibnamefont{and}
  \bibinfo{author}{\bibfnamefont{M.}~\bibnamefont{Lieberman}},
  \emph{\bibinfo{title}{Regular and Stochastic Motion}}
  (\bibinfo{publisher}{Springer-Verlag}, \bibinfo{year}{1982}).

\bibitem[{\citenamefont{Contopoulos}(2002)}]{OrderChaosDynAstro}
\bibinfo{author}{\bibfnamefont{G.}~\bibnamefont{Contopoulos}},
  \emph{\bibinfo{title}{Order and Chaos in Dynamical Astronomy}}
  (\bibinfo{publisher}{Springer}, \bibinfo{year}{2002}), ISBN
  \bibinfo{isbn}{978-3-540-43360-6}.

\bibitem[{\citenamefont{Greene}(1979)}]{Gree79}
\bibinfo{author}{\bibfnamefont{J.~M.} \bibnamefont{Greene}},
  \bibinfo{journal}{J.~Math.~Phys.} \textbf{\bibinfo{volume}{20}},
  \bibinfo{pages}{1183} (\bibinfo{year}{1979}).

\bibitem[{\citenamefont{Giorgilli}(1989)}]{LectureNotesInPhysics}
\bibinfo{author}{\bibfnamefont{A.}~\bibnamefont{Giorgilli}}, in
  \emph{\bibinfo{booktitle}{Integrable Systems and Applications}}, edited by
  \bibinfo{editor}{\bibfnamefont{M.}~\bibnamefont{Balabane}},
  \bibinfo{editor}{\bibfnamefont{P.}~\bibnamefont{Lochak}}, \bibnamefont{and}
  \bibinfo{editor}{\bibfnamefont{C.}~\bibnamefont{Sulem}}
  (\bibinfo{publisher}{Springer Berlin / Heidelberg}, \bibinfo{year}{1989}),
  vol. \bibinfo{volume}{342} of \emph{\bibinfo{series}{Lecture Notes in
  Physics}}, pp. \bibinfo{pages}{142--153}.

\end{thebibliography}

\end{document}